\documentclass[structabstract]{aa}  
\usepackage{graphicx}
\usepackage{amsmath}
\usepackage{txfonts}
\usepackage{subfigure}
\usepackage{natbib}
\bibpunct{(}{)}{;}{a}{}{,}
\usepackage[usenames]{color}

\begin{document}
\frenchspacing
\title{Non-local thermodynamic equilibrium inversions from a 3D MHD chromospheric model} 
\titlerunning{Non-LTE inversions from a 3D MHD chromospheric model}

\author{
  J. de la Cruz Rodr\'iguez\inst{1,2,3},
  H. Socas-Navarro\inst{4},
  M. Carlsson\inst{3,5}
 \and
  J. Leenaarts\inst{3,6} 
}
\authorrunning{J. de la Cruz Rodr\'iguez et al.}

\offprints{J.d.l.C.R. \email{jaime.cruz@physics.uu.se}}

\institute{  
  Department of Physics and Astronomy, Uppsala University, 
  Box 516, SE-75120 Uppsala, Sweden
  \and
  Institute for Solar Physics of the Royal Swedish
  Academy of Sciences, AlbaNova,
  SE-106\,91 Stockholm, Sweden
 \and 
  Institute of Theoretical Astrophysics, University of Oslo, 
  P.O. Box 1029 Blindern, N-0315 Oslo, Norway
  \and
  Instituto de Astrof\'isica de Canarias, Avda V\'ia L\'actea S/N, 
  La Laguna 38205, Tenerife, Spain
  \and
  Center of Mathematics for Applications, University of Oslo, P.O. Box 1053 Blindern, N-0316 Oslo, Norway
 \and
 Utrecht University, P.O. Box 80000  NL--3508 TA Utrecht, The Netherlands
}

\newcommand {\FeI} {\ion{Fe}{i}}
\newcommand {\CaII} {\ion{Ca}{ii}}
\newcommand {\ca} {\ion{Ca}{ii}~$\lambda$8542~\AA\ }
\newcommand {\ltau} {$\log_{10}(\tau_{500})$}
\newcommand {\ms} {m s$^{-1}$}
\newcommand {\kms} {km~s$^{-1}$}
\newcommand {\bz} {$B_z$}
\newcommand {\bx} {$B_x$}
\newcommand {\by} {$B_y$}
\newcommand {\vlos} {$v_{los}$}


\abstract
{The structure of the solar chromosphere is believed to be governed by magnetic fields, even in quiet-Sun regions that have a relatively weak photospheric field. During the past decade inversion methods have emerged as powerful tools for analyzing the chromosphere of active regions. The applicability of inversions to infer the stratification of the physical conditions in a dynamic 3D solar chromosphere has not yet been studied in detail.}
{This study aims to establish the diagnostic capabilities of non-local thermodynamical equilibrium (NLTE) inversion techniques of Stokes profiles induced by the Zeeman effect in the \ca line.}
{We computed the \CaII\ atomic level populations in a snapshot from a 3D radiation-MHD simulation of the quiet solar atmosphere in non-LTE using the 3D radiative transfer code Multi3d. These populations were used to compute synthetic full-Stokes profiles in the \ca line using 1.5D radiative transfer and the inversion code Nicole. The profiles were then spectrally degraded to account for finite filter width and Gaussian noise was added to account for finite photon flux. These profiles were inverted using Nicole and the results were compared with the original model atmosphere.}
{Our NLTE inversions applied to quiet-Sun synthetic observations provide reasonably good estimates of the chromospheric magnetic field, line-of-sight velocities and somewhat less accurate, but still very useful, estimates of the temperature. Three dimensional scattering of photons cause cool pockets in the chromosphere to be invisible in the line profile and consequently they are also not recovered by the inversions. To successfully detect Stokes \emph{linear} polarization in this quiet snapshot, a noise level below 10$^{-3.5}$ is necessary.}
{}

\maketitle
\section{Introduction}\label{introduction}
Measuring the magnetic field of the solar chromosphere is an outstanding challenge for observers. The reason for this is threefold: there are not many lines available in the optical spectrum with a sufficiently high opacity to place the formation in the chromosphere, the polarimetric response is limited, and the few lines available have a non-local component to their formation \citep{1997socas-navarro,2006pietarila, 2010manso, 2012leenaarts}.

The situation is much simpler in the photosphere: to model most \ion{Fe}{i} photospheric lines, it is customary to assume local thermodynamic equilibrium \citep[LTE, see][]{1982rutten}, which means that the atom population densities can be computed directly from the local temperature and electron density at each point. Unfortunately, most chromospheric lines require a non-LTE (NLTE hereafter) treatment of the radiative transfer problem. 

Out of the available chromospheric diagnostic lines, the \ion{Ca}{ii}~infrared (IR) triplet ($\lambda8498, \lambda8542, \lambda8662$) constitutes a good compromise of modeling efforts, polarimetric sensitivity and observational requirements; 
partial redistribution effects are negligible \citep{Uitenbroek1989} and almost the entire calcium is singly ionized under typical solar chromospheric conditions such that time-dependent ionization effects are negligible and the statistical equilibrium equations can be used to compute the level populations  \citep{2011sven}. 

The \ion{Ca}{ii}~IR triplet lines have relatively low effective Land\'e factors ($\mathrm{g}_{8498}=1.07$, $\mathrm{g}_{8542}=1.10$, $\mathrm{g}_{8662}=0.87$) and broad Gaussian line cores compared to photospheric lines. This results in low polarization signals compared to those from photospheric lines. However, the triplet lines have the advantage that they can be observed over the entire solar disk (as opposed to the \ion{He}{i}~$\lambda10830$ multiplet, which is visible only in active patches) and provide information also on the plasma temperature.

The development of inversion techniques has improved our capabilities to infer physical quantities from the chromosphere \citep[see][and references therein]{2010trujillo-bueno,2012asensio-ramos}. The first full-Stokes NLTE inversions of the \ion{Ca}{ii} infrared lines were carried out by \citet{2000socas-navarro}. Their approach is ideal for analyzing active region and network observations where the magnetic field is strong. To analyze quiet-Sun observations, very high sensitivity is required, especially in Stokes~$Q$ and $U$, where signals are expected to be very low.

This study aims to improve our understanding of  chromospheric magnetic field diagnostics. To this end we will use 3D simulations to study NLTE inversions of quiet-Sun observations to quantify their reliability. We will also study the detectability of Zeeman-induced polarization; instrumental limitations and requirements.
The goal is to investigate the polarimetric properties of the \ion{Ca}{ii}~IR lines and the regime in which inversions yield realistic results. 

A 3D MHD simulation is used to compute synthetic full-Stokes profiles with a realistic 3D solution of the NLTE problem. The resulting synthetic profiles are inverted assuming plane-parallel geometry on a pixel-by-pixel basis. Realistic values of noise and instrumental degradation are applied to the simulated observations before carrying out the inversions. 

In real observations, Stokes $Q$ and $U$ are produced by the combined action of scattering polarization and transverse Zeeman effect when $10 \ \text{G}<B<100$ G \citep{2010manso}. Within the mentioned range, Hanle signals are expected to have $Q$ and $U$ amplitudes $\leq 10^{-4}$ at solar disk center and $\leq 10^{-3}$ close to the limb. In active regions the Zeeman contribution would be dominant. We emphasize that this study only accounts for Zeeman-induced polarization, a simplification that is needed given the computational demands of the problem \citep[see][]{2012carlin}. 

The 3D simulation, the radiative transfer calculations and the simulated observations are explained in Sect.~\ref{sec:simobs}, the detectability of the Stokes signal is discussed in Sect.~\ref{sec:observability} and the inversion code and the inversion results are discussed in Sect.~\ref{nicinv}. The conclusions are given in Sect.~\ref{conclusions}.

\section{Simulated observations}
\label{sec:simobs}

\subsection{3D MHD model}\label{model}

We used a snapshot from a 3D radiation-MHD simulation of the quiet-Sun that extends from the upper convection zone to the low corona. The simulation was carried out with the Oslo stagger code \citep[OSC,][]{2007hansteen} which includes non-gray radiative losses using multi-group opacities with scattering, NLTE radiative losses in the chromosphere, optically thin radiative losses in the corona and  thermal conduction along the magnetic field lines.
The simulation domain consists of a grid of $256\times128\times160$ points covering a physical range of $16.6\times8.3\times15.5$ Mm, extending from the upper convection zone to the lower corona (from 1.5 Mm below to 14 Mm above average optical depth unity at 5,000~\AA). The simulation has an average magnetic field strength of 150~G in the photosphere, a value consistent with the mean field strength inferred by \citet{2004trujillo-bueno} via the Hanle effect. The magnetic field is mostly concentrated in intergranular lanes at photospheric heights, whereas in the chromosphere it tends to fill up more space. At $\mathrm{z}=1000$~km, the average magnetic field strength is 79~G.

\begin{figure}[]
      \centering
	\resizebox{\hsize}{!}{\includegraphics[trim = 0 0 0 0, clip]{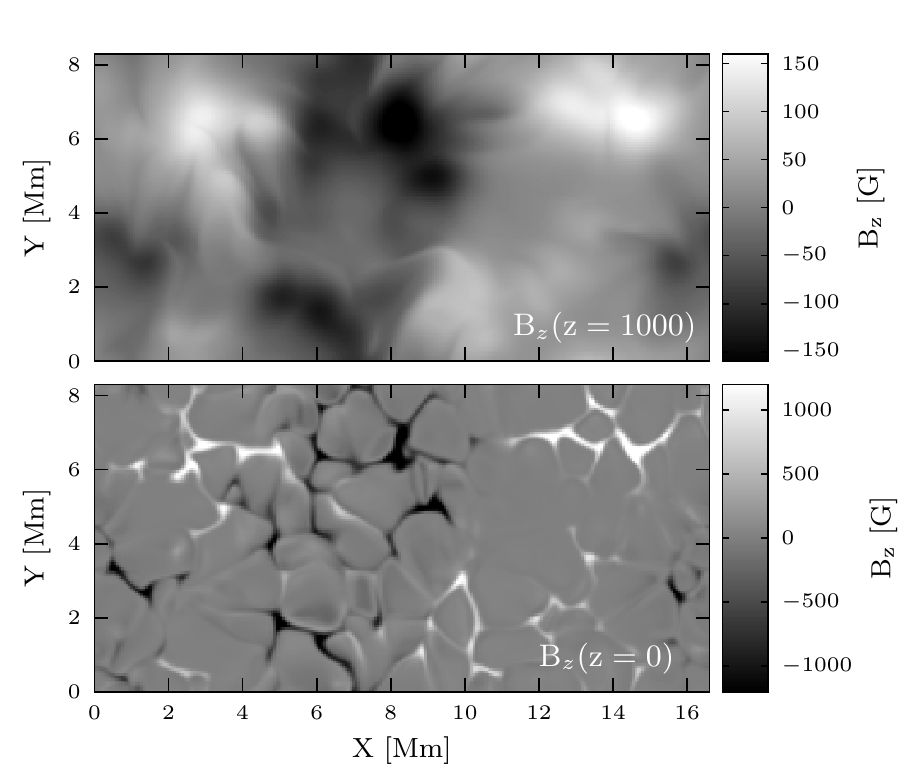}}	
        \caption{Vertical component of the magnetic field, $B_z$, at horizontal slices through the 3D model atmosphere at $z$=1000 km (top) and $z$= 0 km (bottom).}
        \label{mslice}
\end{figure}

Fig.~\ref{mslice} illustrates the vertical component of \vec{B} in the photosphere (bottom) and in the low chromosphere (top). The simulation snapshot used in this work is the same as in \citet{2009leenaarts,2010leenaart1} and \citet{2011rutten1}. \\

\subsection{Radiative transfer}\label{nicole}

Our synthetic full Stokes profiles are computed in two steps. First, population densities are calculated with  \textsc{Multi3D}  \citep{2009leenaarts1},  which allows one to evaluate the three-dimensional radiation field (3D NLTE hereafter). The \ion{Ca}{ii} atom model used in this work consists of five bound levels plus a continuum. We calculate the departure coefficients, defined as
\begin{equation}
	b_i = \frac{n_i}{n_i^*} \label{depcoef},
\end{equation}
where $n_i$ is the population density of an atomic level $i$ computed in NLTE and $n_i^*$ is the equivalent LTE population density.

In the second step, the same atmospheric model is used to compute the LTE solution with the non-LTE inversion code based on the Lorien engine \citep[\textsc{Nicole,}][]{nicoleref}. 
The departure coefficients obtained previously with \textsc{Multi3D} are applied to the LTE populations. Then, each column is resampled to a grid that is optimized for the radiative transfer computations in \CaII\ lines: for each column the model is truncated at the height where the temperature is higher than 50,000~K. The grid points are then distributed according to gradients in temperature, density and opacity, placing more points where the gradients are steep. The temperature, electron density, mass density, velocity, and departure coefficients are interpolated to the new grid using an accurate interpolation algorithm based on Hermitian splines \citep{1982intep}. 

\citet{2009leenaarts} pointed out that the Gaussian core of the synthetic \ca Stokes $I$ profile is narrower and \emph{steeper} than those from observations, probably from the lack of small scale random motions in the model. Assuming Zeeman splitting in a weak-field regime, the shape of Stokes $Q$ and $V$ can be expressed as a function of the first derivative of the intensity \citep[][]{2004landi}.
\begin{eqnarray}
	Q\left(\lambda\right) &=& -\frac{1}{4}\Delta\lambda^2_B \bar{G}\frac{\eta ''}{\eta '} \left( \frac{\partial I}{\partial \lambda} \right)\label{wq}  \\
	V\left(\lambda\right) &=& -\Delta\lambda_B \bar{g} \cos \theta \left( \frac{\partial I}{\partial \lambda} \right)\label{wv},
\end{eqnarray}
where $\Delta\lambda_B$ represents the Zeeman splitting of the line, $\bar{g}$ is the Land\'e factor, $\theta$ is the inclination, $\bar{G}$ depends on the quantum numbers of the line and $\eta$ is the Voigt profile. According to Eqs.~\ref{wq}~and~\ref{wv} the amplitude of Stokes~$Q$ and $V$ is enhanced if the slope of Stokes $I$ is steep. Thus, steeper Stokes $I$ profiles produce more strongly peaked Stokes $Q$, $U$ and $V$ profiles. Therefore, our Stokes $Q$, $U$ and $V$ spectra would be unrealistically strong and narrow, which would lead to an overestimation of the effect of spectral smearing and an underestimation of the effect of photon noise. 

\begin{figure}[]
      \centering
        \resizebox{\hsize}{!}{\includegraphics[trim=0cm 0.0cm 0cm 0cm, clip]{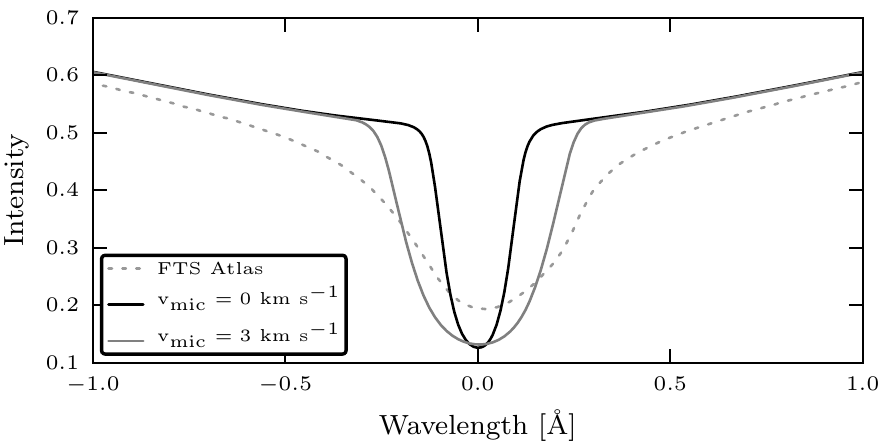}}
        \caption{Spatially-averaged profile without micro-turbulence (black-solid), with micro-turbulence (gray-solid) and the FTS atlas (gray-dashed).}
        \label{vmic}
\end{figure}

To make more realistic predictions, we introduced an artificial micro-turbulence in the model. Without micro-turbulence, the full width at half maximum (FWHM) of the chromospheric core, measured in our spatially resolved spectra is about 180~m\AA. Corresponding measurements from quiet-Sun line profiles were carried out by \citet{2009cauzzi}, who found values in the range 450-550 m\AA. This means that our profiles are of the order of a factor 2.5 narrower than observed, assuming that instrumental degradation has a negligible effect on the FWHM of the profiles. By artificially introducing micro-turbulence of 3~\kms \ the width of our profiles is increased up to $\sim 400$ m\AA. 
A comparison between the spatially averaged spectrum with and without micro-turbulence is given in Fig.~\ref{vmic}. 

After adding the micro-turbulence the formal solution of the polarized radiative transfer equation was computed with \textsc{Nicole} using a Hermitian method, developed by \citet{1998bellot-rubio}. This two-step process from computing the Stokes profiles is necessary because \textsc{Multi3D} cannot compute polarized radiative transfer, and \textsc{Nicole} cannot compute 3D radiative transfer. Only by combining the two codes we obtain full Stokes profiles computed with 3D radiative transfer that serve as our synthetic observations for the inversion.

\begin{figure*}[]
	\centering

\resizebox{0.9\hsize}{!}{\includegraphics[]{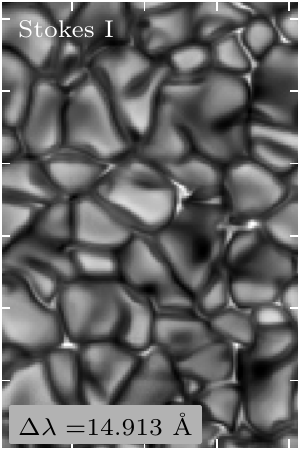}\includegraphics[]{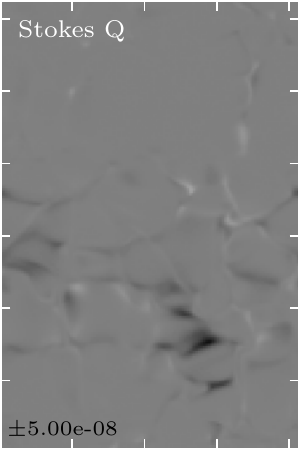}\includegraphics[]{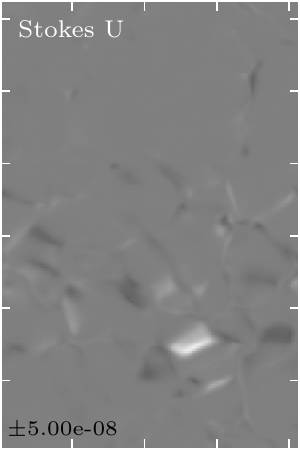}\includegraphics[]{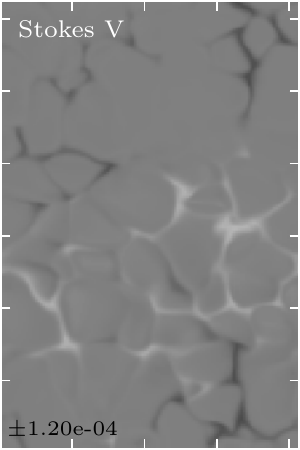}\includegraphics[]{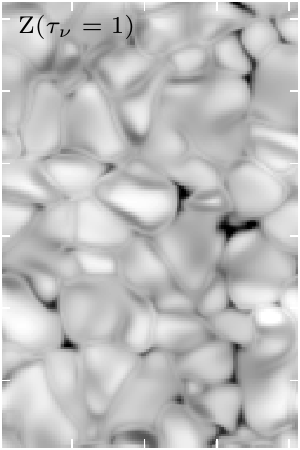}\includegraphics[]{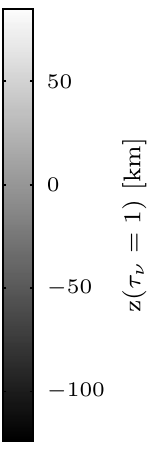}}\vspace{-0.033cm}
\resizebox{0.9\hsize}{!}{\includegraphics[]{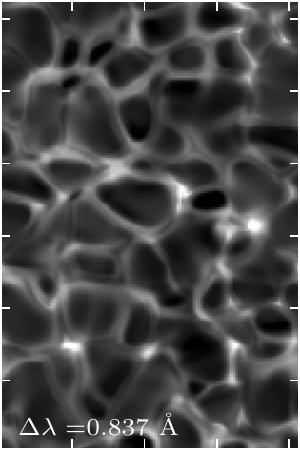}\includegraphics[]{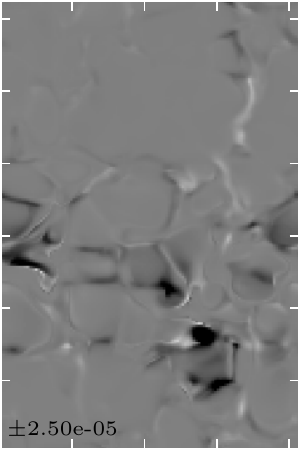}\includegraphics[]{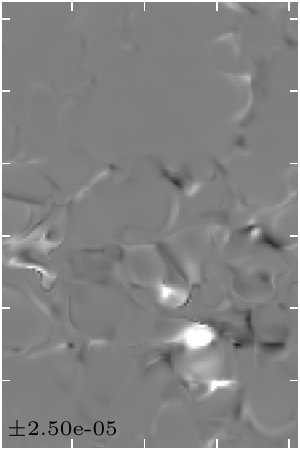}\includegraphics[]{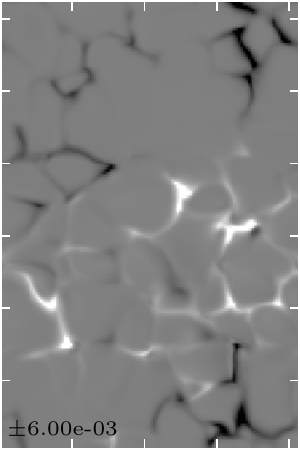}\includegraphics[]{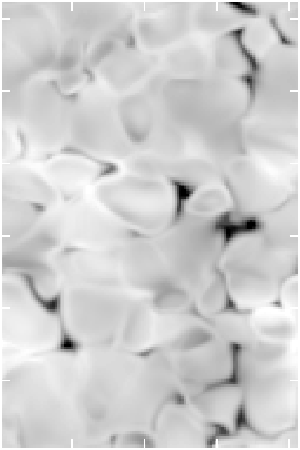}\includegraphics[]{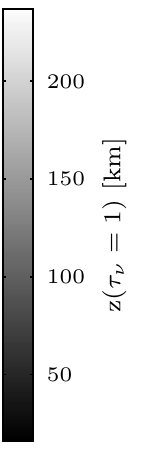}}\vspace{-0.033cm}
\resizebox{0.9\hsize}{!}{\includegraphics[]{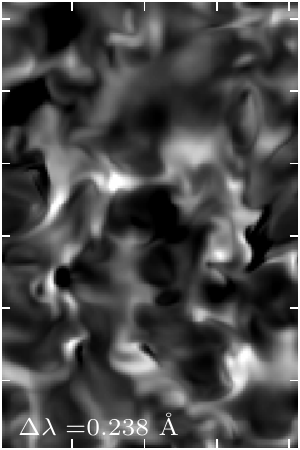}\includegraphics[]{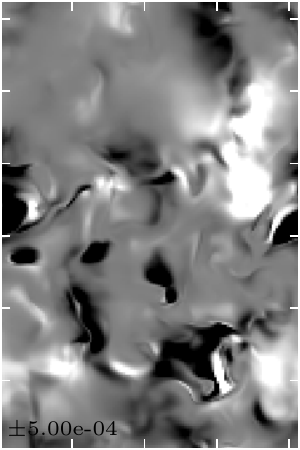}\includegraphics[]{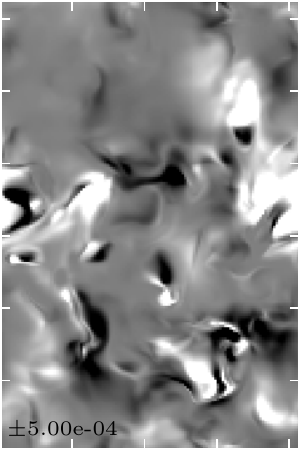}\includegraphics[]{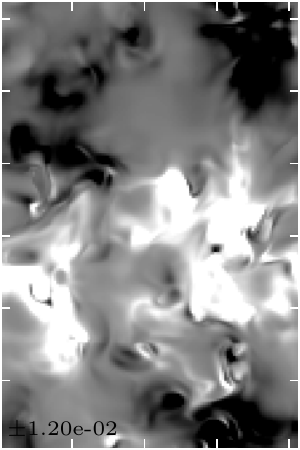}\includegraphics[]{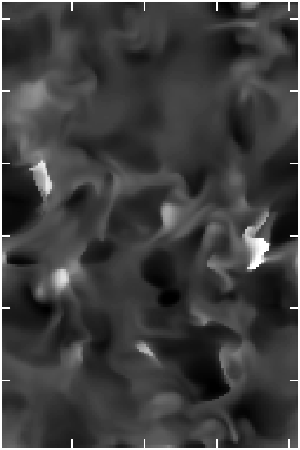}\includegraphics[]{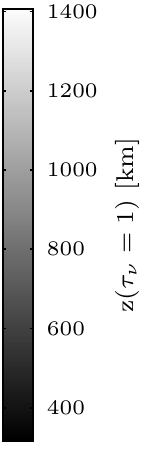}}\vspace{-0.033cm}
\resizebox{0.9\hsize}{!}{\includegraphics[]{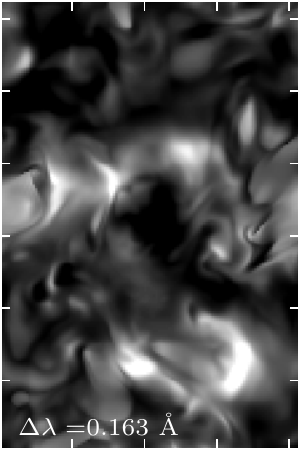}\includegraphics[]{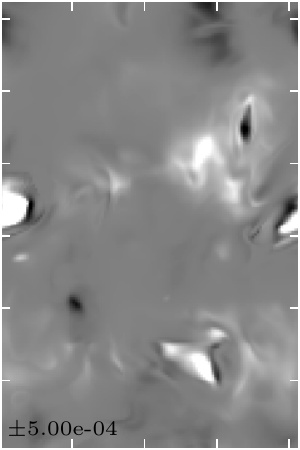}\includegraphics[]{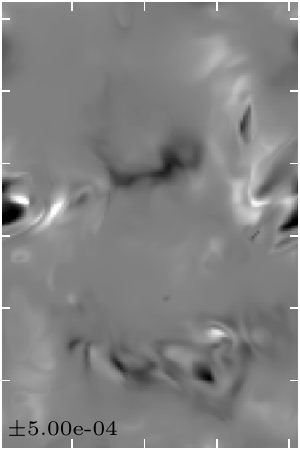}\includegraphics[]{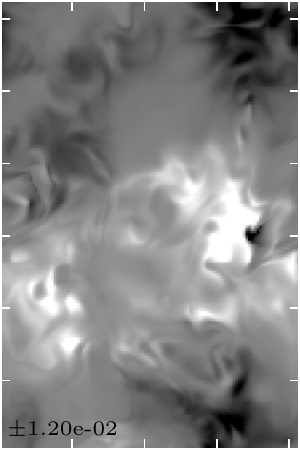}\includegraphics[]{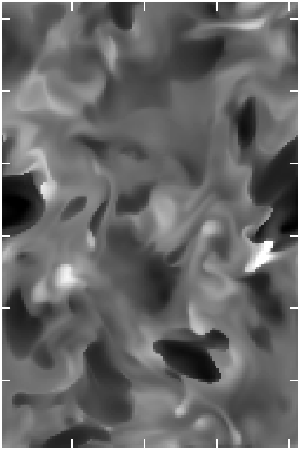}\includegraphics[]{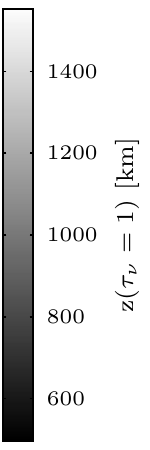}}\vspace{-0.033cm}
\resizebox{0.9\hsize}{!}{\includegraphics[]{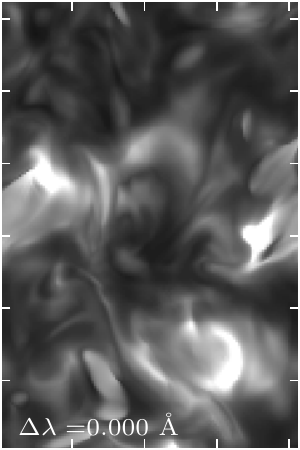}\includegraphics[]{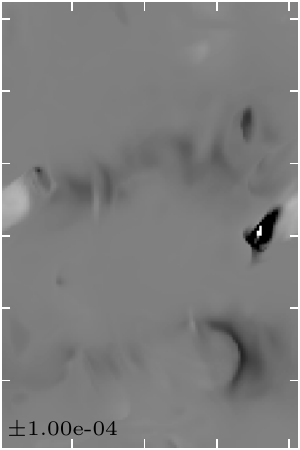}\includegraphics[]{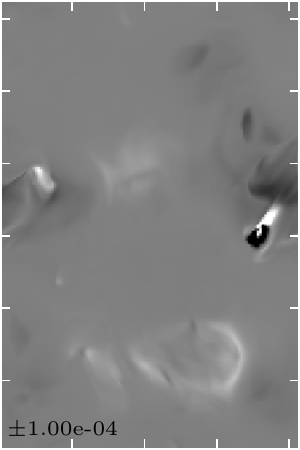}\includegraphics[]{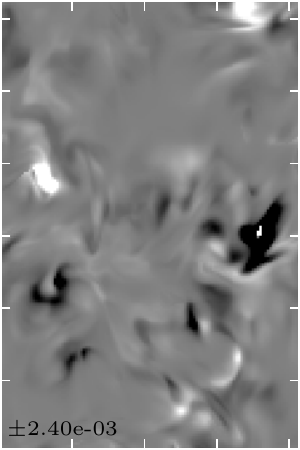}\includegraphics[]{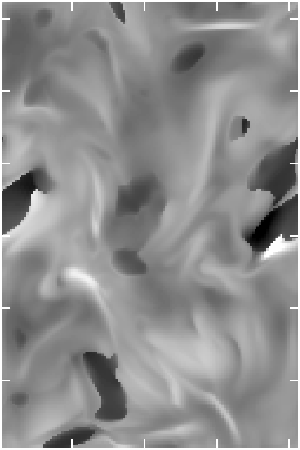}\includegraphics[]{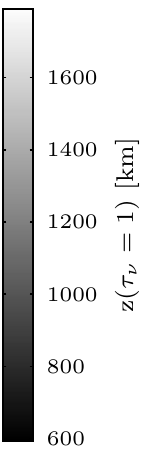}}\vspace{-0.033cm}
        \caption{Synthetic full-Stokes images computed from a $12.45\times8.3$ Mm patch of the model. From left to right, the first 4 columns show monochromatic Stokes $I$, $Q$, $U$ and $V$ images, respectively. The last column shows the height where the monochromatic optical depth equals one. From top to bottom, the wavelength decreases from the far red wing to the core of the line. The wavelength relative to line core is indicated on each of the Stokes $I$ panels.  The Stokes $Q$, $U$ and $V$ panels have been scaled independently to enhance visibility of weak polarization features; the scale range is indicated within each panel in units of the continuum intensity. The tick-mark separation is 2 Mm.  The panels have been transposed with respect to the panel orientation of Fig. 1.}\label{fig:scan_stokes}
\end{figure*}

Our synthetic intensity profiles are near-identical to the results of \citet{2009leenaarts}, with small remaining differences due to small differences in the radiative transfer codes.
In Fig.~\ref{fig:scan_stokes}, monochromatic images at $\Delta\lambda=14.9$, $0.84$, $0.24$, $0.16$ and $0.00$ \AA \ from the core of the line illustrate the vast formation range of the \ion{Ca}{ii}~$\lambda8542$ \AA \ line, which covers the photosphere and part of the chromosphere. The polarization profiles are induced by Zeeman splitting. The Stokes~$Q$ and $U$ profiles show amplitudes about $10^{-4}$ (relative to the continuum intensity), peaking around $10^{-3}$ at locations where the magnetic field is (relatively) strong and mostly horizontally oriented. The Stokes $V$ signal is one order of magnitude stronger than Stokes $Q$ and  $U$, usually within the range $[10^{-3}, 10^{-2}]$, and shows extended areas with the same polarity. 

The last column in Fig.~\ref{fig:scan_stokes} illustrates for each wavelength the height where the monochromatic optical depth equals one, $Z(\tau_\nu$=1). In the wings of the line, the height of the $\tau_\nu$=1 surface is dominated by the temperature sensitivity of the opacity. This gives rise to granular "hills" with a higher $\tau_\nu$=1 surface than in the cooler intergranular lanes. Additionally, plasma evacuation in magnetic elements reduces the opacity, allowing to \emph{see} deeper into the atmosphere \citep[e.g.][]{2004keller}. At 
$\Delta\lambda$=0.837~{\AA} we sample the inverse granulation layer where the temperature contrast is lower and most of the variation in the height of the $\tau_\nu$=1 surface is caused by the lower plasma density where the magnetic field is strong. Closer to line center there is a strong variation in the $\tau_\nu$=1 surface caused by strong temperature and density variations and Doppler shifts of the line core into or out of 
the passband.

\section{Observability of Stokes signals}\label{sec:observability}

This study is partially motivated by the instrumental requirements of future telescopes. In this context, Fabry-P\'erot Interferometers (FPI) have become very popular because they allow a balanced trade-off between cadence, spatial resolution and spectral resolution. 
We analyze here the combined effects of limited spectral resolution and noise in our synthetic observations.

We assumed a Gaussian spectral filter transmission and additive photon noise produced by the finite photon flux. The instrumental point spread function (PSF) is characterized by its FWHM, and hereafter we refer to this parameter when instrumental spectral smearing is discussed. 
The additive noise is characterized in terms of its standard deviation $\sigma$ relative to the continuum intensity.

\begin{figure*}[]
      \centering   
\resizebox{\hsize}{!}{\includegraphics[]{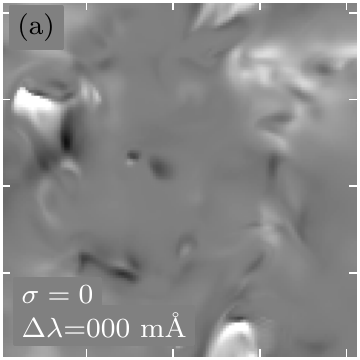}\includegraphics[]{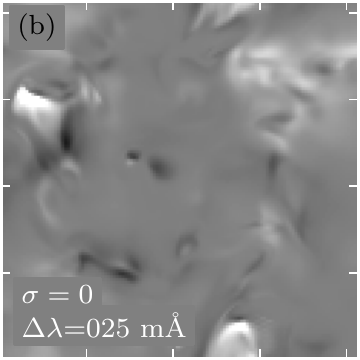}\includegraphics[]{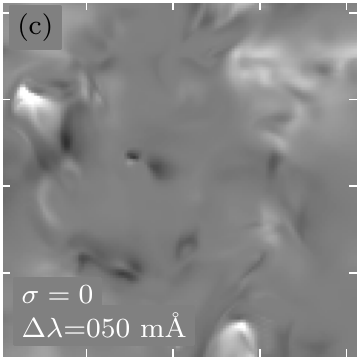}\includegraphics[]{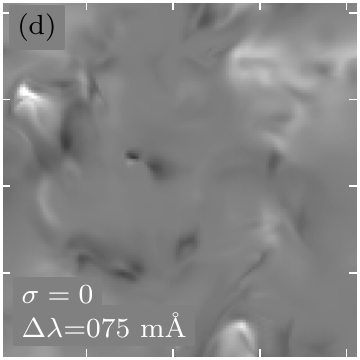}\includegraphics[]{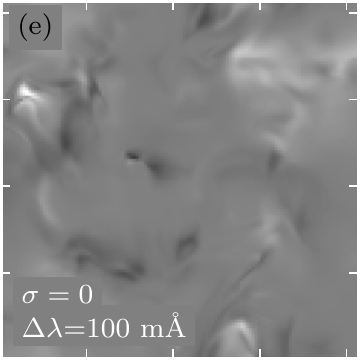}}\vspace{-0.033cm}
\resizebox{\hsize}{!}{\includegraphics[]{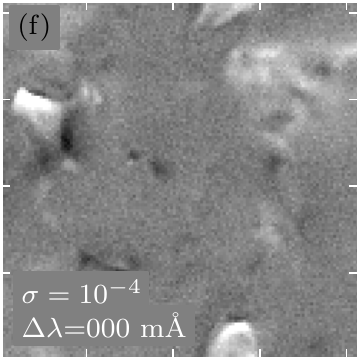}\includegraphics[]{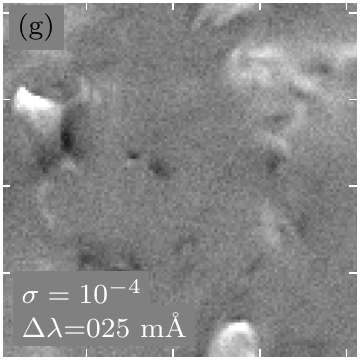}\includegraphics[]{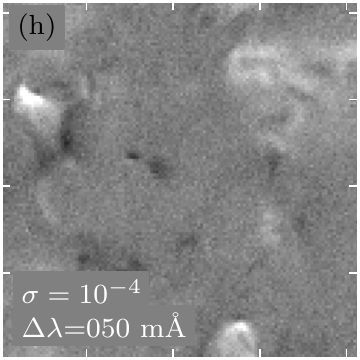}\includegraphics[]{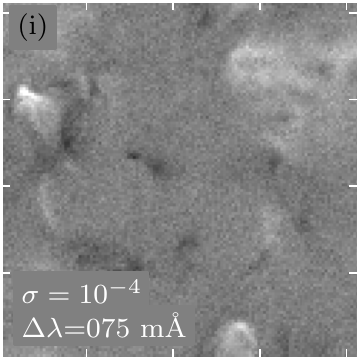}\includegraphics[]{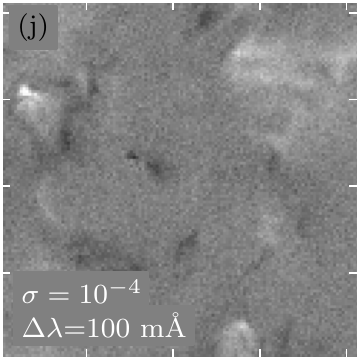}}\vspace{-0.033cm}
\resizebox{\hsize}{!}{\includegraphics[]{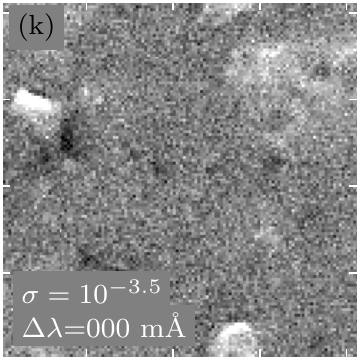}\includegraphics[]{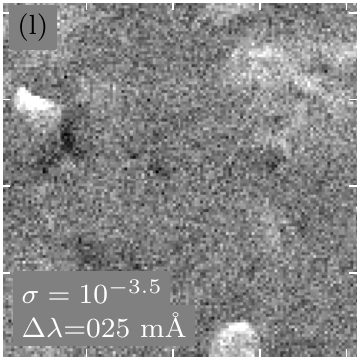}\includegraphics[]{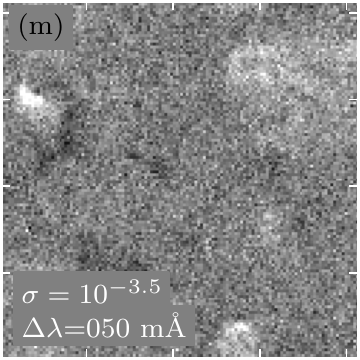}\includegraphics[]{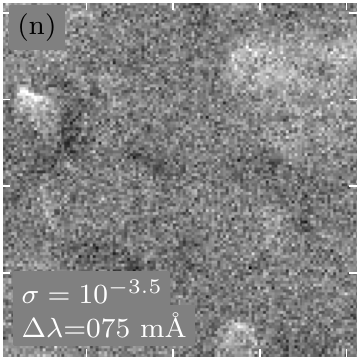}\includegraphics[]{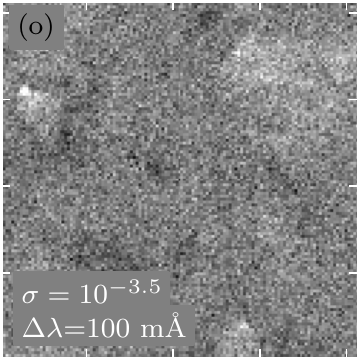}}\vspace{-0.033cm}
\resizebox{\hsize}{!}{\includegraphics[]{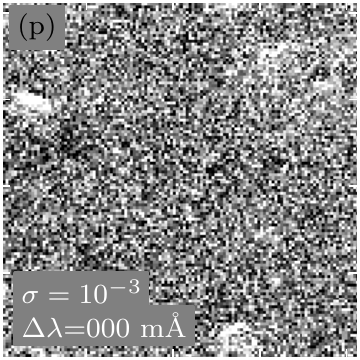}\includegraphics[]{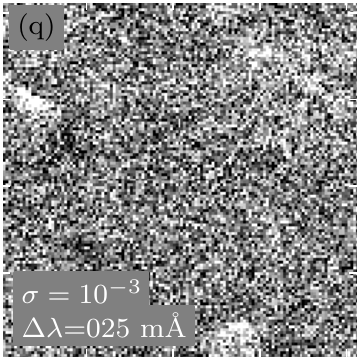}\includegraphics[]{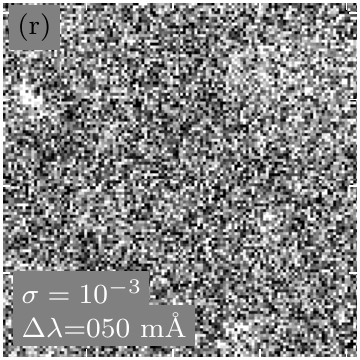}\includegraphics[]{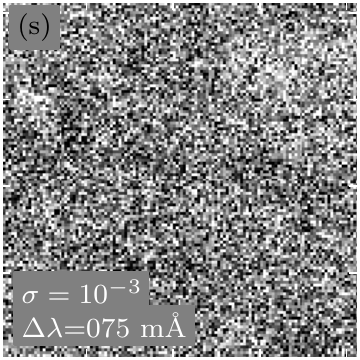}\includegraphics[]{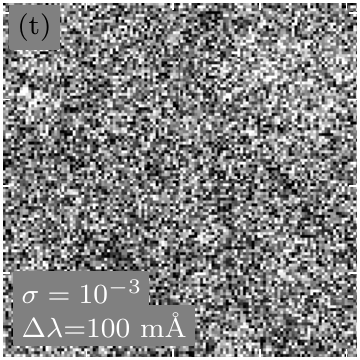}}\vspace{-0.033cm}
        \caption{Stokes $Q$ image spectrally degraded with Gaussian PSFs and random noise. From left to right, the spectral degradation increases  as a function of the FWHM of the Gaussian, from $\Delta \lambda=0$ (no spectral smearing) to  $\Delta \lambda=100$ m\AA. From top to bottom, the noise component increases, from $\sigma=0$ to $\sigma=10^{-3}$. The images are scaled to $\pm 0.15\%$ relative to the continuum intensity. }
        \label{dstokesQ}
\end{figure*}

\begin{figure*}[]
      \centering	\resizebox{\hsize}{!}{\includegraphics[]{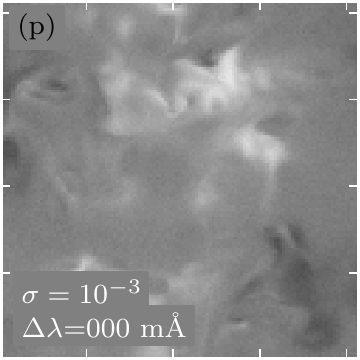}\includegraphics[]{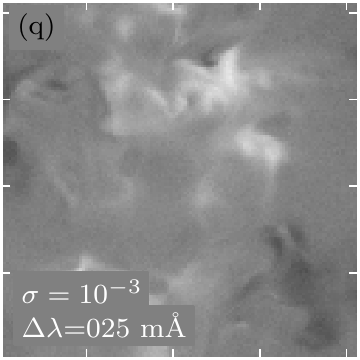}\includegraphics[]{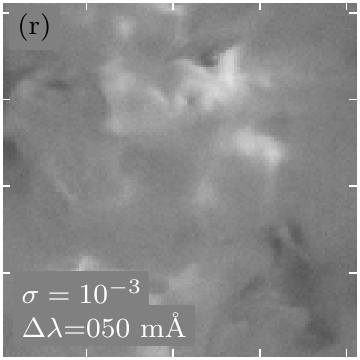}\includegraphics[]{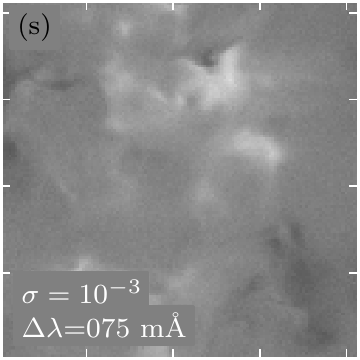}\includegraphics[]{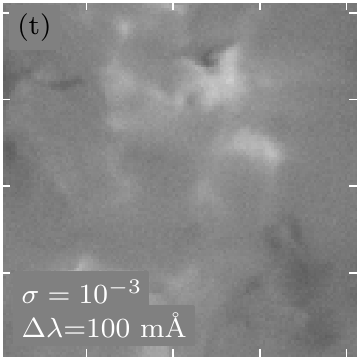}}
	        \caption{Stokes $V$ image spectrally degraded with Gaussian PSFs. From left to right, the spectral degradation increases as a function of the FWHM of the Gaussian, from $\Delta \lambda=0$ (no spectral smearing) to  $\Delta \lambda=100$ m\AA. A Gaussian random noise component of $\sigma=10^{-3}$ has been added. The images are scaled to $\pm 5\%$ relative to the continuum intensity.}
        \label{dstokesV}
\end{figure*}

Our results are illustrated in Figs.~\ref{dstokesQ}~and~\ref{dstokesV} at +205~m\AA \ from the line core, where the red peak of polarization is, on average, located. 

In Fig.~\ref{dstokesQ} we show the Stokes $Q$ signal in a subfield of the synthetic image for various combinations of spectral degradation and noise. The spectral degradation spans the range $[0,100]$ ~m\AA, in 25~m\AA \ steps, the noise levels are $\sigma=0$, $10^{-4}$, $10^{-3.5}$ and $10^{-3}$. The noise value and spectral degradation are indicated inside each panel. 

In the top row of Fig.~\ref{dstokesQ} only spectral degradation is applied. Even though the amplitude of Stokes~$Q$ is decreased by a factor 2 between $\mbox{FWHM}=0$~m\AA \ and $\mbox{FWHM}=100$~m\AA, the main features are still visible at maximum spectral degradation.

The middle rows (panels \emph{(f) - (o)}) correspond to $\sigma=10^{-4}$ and $\sigma=10^{-3.5}$. Stokes~$Q$ features are still above the noise for all assumed values of the FWHM. This result is somewhat expected because our Stokes~$Q$ profiles usually peak within $[10^{-4}, 10^{-3}]$, so noise  only affects the very weak polarization signals. 

The bottom row of Fig. \ref{dstokesQ} corresponds to $\sigma=10^{-3}$, where noise clearly dominates the Stokes $Q$ images. Some regions with strong polarization in $Q$ are discernible in panels \emph{(p)} and \emph{(q)} ($\mbox{FWHM}~\le~25$~m\AA) but are barely visible when $\mbox{FWHM}~=~50$~m\AA. Above this value, Stokes $Q$ vanishes below the noise because of spectral degradation.

Fig. \ref{dstokesV} shows Stokes~$V$ images for the same subfield as Fig.~\ref{dstokesQ}, for the same values of spectral degradation. Only the case corresponding to $\sigma=10^{-3}$ is shown because the signal is well above the noise in all other cases.

Above, we have described and discussed several combinations of noise and spectral resolution. However, for a given telescope aperture, the signal-to-noise ratio can be derived as a function of the instrumental efficiency, pixel size and exposure time. To estimate the number of photons that can be detected at the core of the \ca line, we computed the continuum intensity at 8542~\AA \ using \textsc{Multi} \citep[see][]{1986carlsson} and the FALC atmospheric model by \citet{2006fontenla}. The intensity ratio between the continuum and the core of the line ($\mathrm{I}_0$) is derived from the Fourier Transform Spectrometer at the McMath-Pierce Telescope (hereafter called the FTS atlas) of \citet{fts-atlas}, resulting in $\mathrm{I}_0=\mathrm{I}_\mathrm{lc} / \mathrm{I}_\mathrm{cont}~=~0.18$. For a telescope aperture of 1.5 m, a spectral resolution of $\mathrm{R}~=~200000$ (42 m\AA \ at $\lambda 8542$), a pixel size of $0.1\arcsec$ and an overall system efficiency of 0.1 (telecope-instrument-detector), one can detect  $\mathrm{n}_0 =2.66 \times 10^6$ photons per second. This corresponds to a photon noise level of $6.1 \times 10^{-4}$. 

\begin{figure*}[]
      \centering
      \resizebox{0.80\hsize}{!}{\includegraphics[]{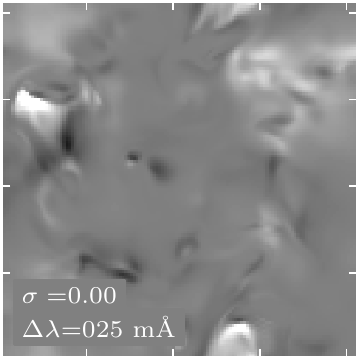}\includegraphics[]{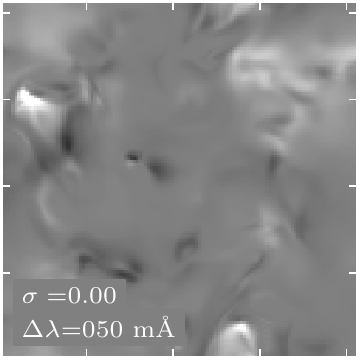}\includegraphics[]{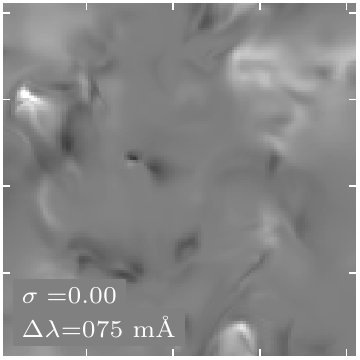}\includegraphics[]{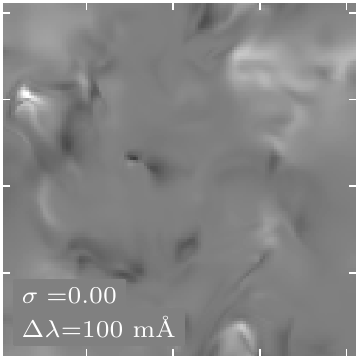}}\vspace{-0.032cm}
\resizebox{0.80\hsize}{!}{\includegraphics[]{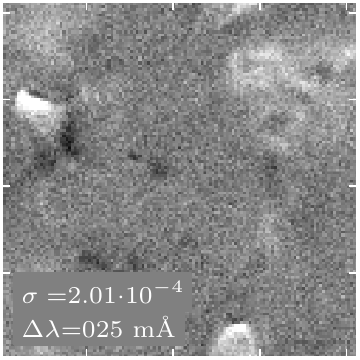}\includegraphics[]{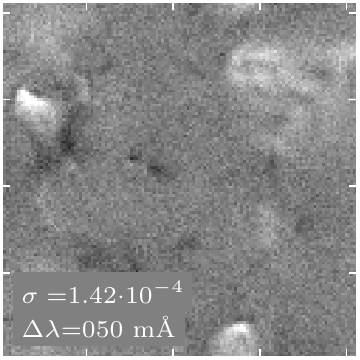}\includegraphics[]{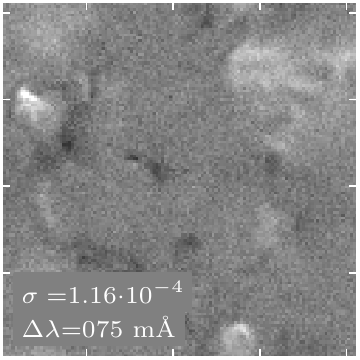}\includegraphics[]{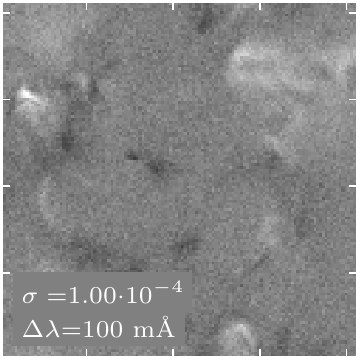}}\vspace{-0.032cm}
\resizebox{0.80\hsize}{!}{\includegraphics[]{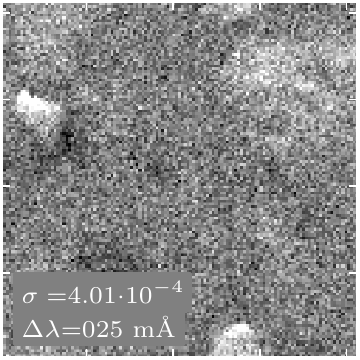}\includegraphics[]{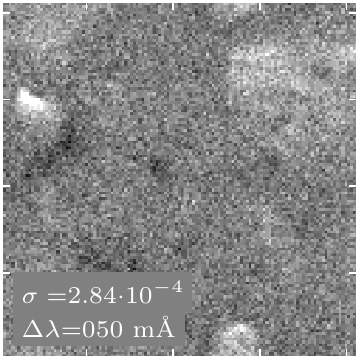}\includegraphics[]{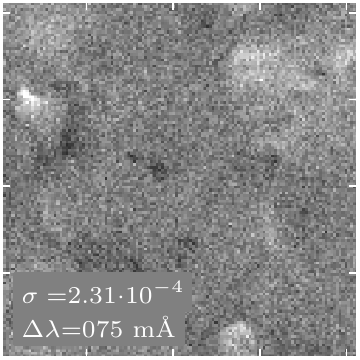}\includegraphics[]{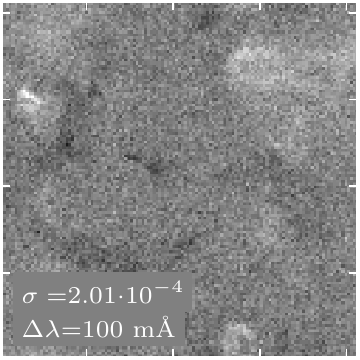}}\vspace{-0.032cm}
\resizebox{0.80\hsize}{!}{\includegraphics[]{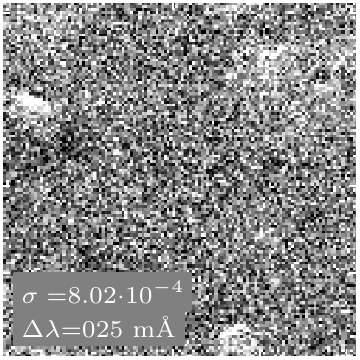}\includegraphics[]{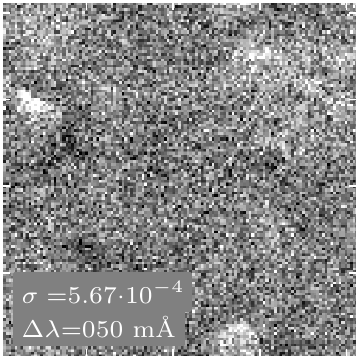}\includegraphics[]{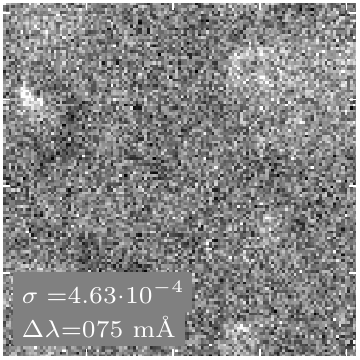}\includegraphics[]{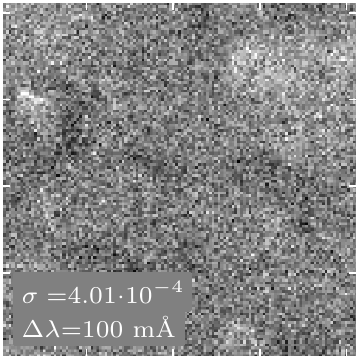}}\vspace{-0.032cm}
        \caption{Stokes $Q$ image spectrally degraded with Gaussian PSFs and random noise. From left to right, the spectral degradation increases  as a function of the FWHM of the Gaussian, from $\Delta \lambda=25$ m\AA \   to  $\Delta \lambda=100$ m\AA. From top to bottom, the integration time is $\infty$, 16 s, 4 s and 1 s, respectively. The images are scaled to $\pm 0.15\%$ relative to the continuum intensity. For a fixed exposure time the noise in each image increases with decreasing width of the instrumental profile. }
        \label{fig-stokesQ-obs}
\end{figure*}

Fig.~\ref{fig-stokesQ-obs} illustrates similar results for several combinations of exposure times ($\infty$ s, 16 s, 4 s and 1 s) and spectral degradation (0~m\AA, 25~m\AA, 50~m\AA \ and  100~m\AA). One can see the difficulties that arise from having a too high spectral resolution: as the instrumental transmission profile becomes narrower, polarization features become more prominent but the noise is also significantly higher. By increasing the exposure time, noise is decreased, but as we explain below, image degradation can appear from the rapid chromospheric motions. Additionally, higher spectral resolution increases the number of wavelength points that are needed to sample the line profile. For clarity, Table~\ref{table-noise} summarizes the numbers used in Fig.~\ref{fig-stokesQ-obs}.

\begin{table}
  \centering
  \caption{\label{table-noise} Photon noise levels as a function of the exposure time and the spectral resolution.} 
  \begin{tabular}{l | c c c c}
    \hline\hline
    \  & 25 m\AA\  & 50 m\AA \  & 75 m\AA \  & 100 m\AA  \\
    \hline
    $16$ s & $2.00\times 10^{-4}$ &$1.42\times 10^{-4}$ &$1.16\times 10^{-4}$& $1.00\times 10^{-4}$ \\
    $4$ s & $4.00\times 10^{-4}$ &$2.83\times 10^{-4}$ &$2.31\times 10^{-4}$& $2.00\times 10^{-4}$ \\
    $1$ s& $8.02\times 10^{-4}$ &$5.67\times 10^{-4}$ &$4.63\times 10^{-4}$& $4.00\times 10^{-4}$ \\
    \hline
  \end{tabular}
\end{table}

The FPI instruments that are currently operating in solar telescopes can typically achieve a spectral resolution of 45-100~m\AA \ in the near-infrared, with noise levels of about $10^{-3}$. Those cases correspond to panels \emph{(r)}, \emph{(s)} and \emph{(t)} in Fig.~\ref{dstokesQ} and \ref{dstokesV}. Our results suggest that Stokes $V$ detections are possible with current instrumentation. However, the detection of linear polarization in the \emph{quiet} Sun requires sensitivities of at least $10^{-3.5}$.

Exposure time in solar observations is limited by the evolution time of the Sun. Longer exposures than the evolution time scale can produce smearing of the data. Particularly, image reconstruction assumes that the Sun is static for a given set of short exposure images. The vigorously dynamic chromosphere shows features moving horizontally within one second \citep{2006noort}. Additionally, FPI instruments can only observe one wavelength at the time, a limiting factor for the wavelength coverage given that all acquisitions within one line scan should be as simultaneous as possible. Therefore observing the chromosphere involves a difficult trade-off between wavelength coverage and exposure time. 

There are ways to ameliorate the sampling. The broad \ion{Ca}{ii} infrared lines have a wide formation range: the fast-moving features observed in the cores contrast with the slow evolution of photospheric granulation present in the wings of the lines. Additionally, the amplitude of Stokes $Q$, $U$ and $V$ rapidly decreases toward the wings of the line. 

\begin{figure}[]
      \centering
      \resizebox{\hsize}{!}{\includegraphics[trim=0cm     0.83cm 0.0cm 0.cm, clip=true]{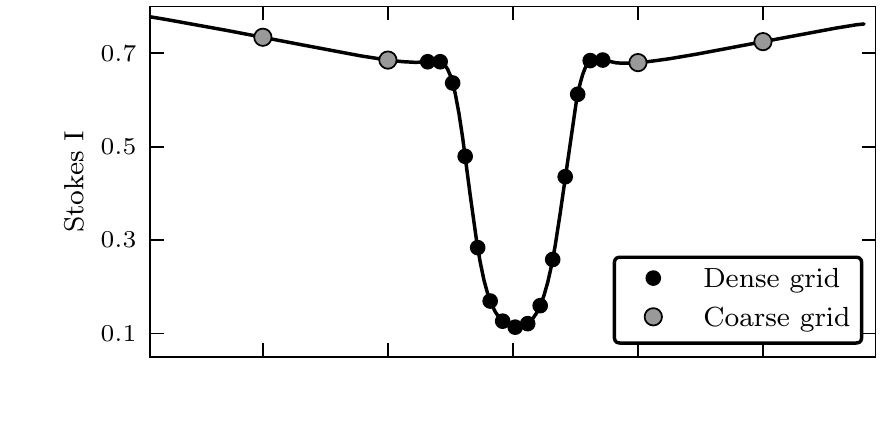}}
        \resizebox{\hsize}{!}{\includegraphics[trim=0cm     0.cm 0.0cm 0.0cm, clip=true]{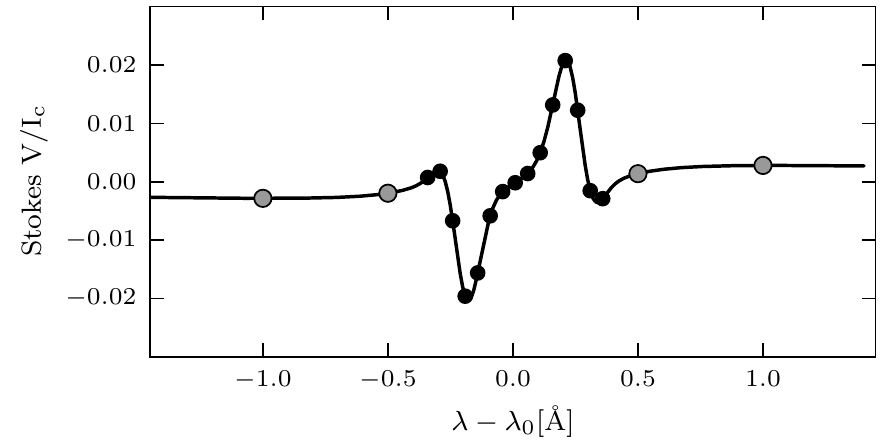}}
        \caption{Stokes $I$ and $V$ profiles from one of the columns from the 3D model. The instrumental spectral FWHM is 100~m\AA. The gray circles indicate a coarse grid of points located outside of the chromospheric range of the line. The black circles sample the core of the line every 50~m\AA.}
        \label{sampling}
\end{figure}

A strategy to optimize the polarization measurements and achieve proper time sampling would be to place a fine grid of points in the chromospheric core and a coarser grid in the wings, where slower variations are expected. Because the photosphere evolves on a longer time scale than the chromosphere, exposures in the wings do not have to be recorded as frequently as in the core. Fig.~\ref{sampling} illustrates this scheme, with a fine grid of points (black circles) distributed in the 
core every 50 m\AA \ and a coarser grid in the wings (gray circles) placed every 500 m\AA. 

\section{Inversions}\label{nicinv}

We used the code \textsc{Nicole} for the inversions presented in this section.
It solves the NLTE problem assuming plane-parallel geometry, isotropic scattering and complete frequency redistribution, using the strategy described by  \citet{1997socas-navarro}. All Zeeman sublevels originating from a given atomic level are assumed to be equally populated, discarding any quantum interference between them, as proposed by \citet{1996trujillo-bueno}. 

The \emph{velocity-free} approximation is used for our calculations here, which assumes a static atmosphere when the atomic-populations are computed. Under these conditions, only half of the profile needs to be computed and fewer quadrature angles are needed. Once the NLTE populations are calculated, the velocity stratification is taken into account to produce the final emerging profiles.

The inversions are initialized with a guess model that contains temperature, electron pressure, line-of-sight (l.o.s.) velocity, micro-turbulence and the three components of the magnetic field vector. The inversion engine of \textsc{Nicole} is based on a Levenberg-Marquardt algorithm that computes the corrections to a guess model in order to minimize the differences between the observed and the synthetic profiles. To maintain physical consistency between temperature, electron pressure and gas pressure, the model is set to hydrostatic equilibrium after each correction. For a given temperature, the gas pressure and the electron pressure are obtained using an equation of state and the hydrostatic equilibrium approximation.

The corrections to the guess model were applied at node points that are equidistantly distributed over the depth scale of the model. The first inversion was initialized with the VAL-C model. After the first inversion another five inversions were performed starting from the VAL-C model where the atmospheric quantities have been given a random offset. This minimized the odds of the algorithm settling into a local minimum of the $\chi^2$ hypersurface.

\begin{figure}[]
  \centering
  \resizebox{\hsize}{!}{\includegraphics[]{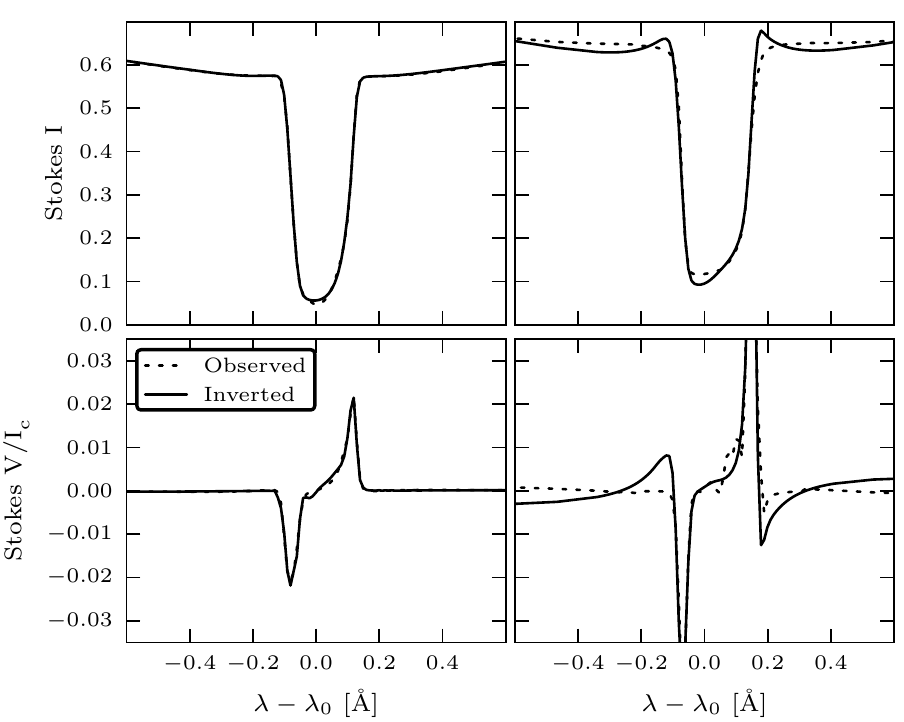}}
  \caption{Illustrative examples of the fits (solid line) to the
    simulated observations (dashed line) at two different locations in
    the LTE inversion of Fig.~\ref{lteres}. Left panels: A good fit,
    at pixel coordinates (25,23) in Fig.~\ref{lteres}. Right panels: A
  comparatively poorer fit at pixel coordinates (39,34). Upper row:
  Stokes~$I$ (normalized to the average continuum intensity). Lower row:
Stokes{$V$}.}
  \label{fig:fitslte}
\end{figure}

Fig.~\ref{fig:fitslte} shows the
difference in the fit between a pixel that converged to the correct
solution and another one for which the algorithm settled in a
secondary minimum of the $\chi^2$ hypersurface. The noise created in
the resulting image by imperfect fits (we will refer to this as {\em
  inversion noise} to avoid confusion with the usual photon noise
found in the observations) can be minimized by implementing more
sophisticated schemes with multiple initializations to ensure that one
reaches the absolute minimum. Genetic algorithms
\citep{1995charbonneau} provide a suitable approach to the problem and
have been successfully implemented in Stokes inversion codes (e.g.,
\citealt{2004lagg}). However, such an algorithm dramatically increases the already very large amount of computational work required for the solution of the NLTE inversion problem. For this reason we have refrained from implementing anything
beyond the already mentioned five attempts with randomized
initializations. 

The spectral resolution of the instrument, the wavelength coverage and sampling and the signal-to-noise ratio set constraints on the amount of information that can be retrieved by an inversion. For a given set of these quantities there is a maximum number of nodes after which adding more nodes does not improve the inversion of the data. Adding more nodes can lead instead to artifacts. 

To improve convergence, the inversions were computed in two cycles \citep[see][]{1992ruiz-cobo,2007pietarila, 2011socas-navarro}. In the first cycle, a limited set of nodes was used to obtain a rough solution that was then refined during the second cycle with more nodes. A summary of the number of nodes used on each cycle is shown in Table~\ref{table:nodes}.

\begin{table}
  \centering
  \caption{\label{table:nodes} Summary of node points used during each cycle of the inversion.} 
  \begin{tabular}{ r | c c}
    \hline\hline
    Physical parameter & Nodes in cycle 1 & Nodes in cycle 2 \\   
    \hline
    Temperature & 6 & 10\\
    l.o.s velocity & 2 & 5 \\
    $B_z$          & 1 & 4 \\
    $B_x$          & 0 & 1 \\
    $B_y$          & 0 & 1 \\
    \hline
  \end{tabular}
\end{table}

The four Stokes parameters are weighted differently in the inversion. The following ratio was used in the present work: $\mathrm{W}(I,Q,U,V) = [1,20,20,8]$.

The inversions were computed with critically sampled profiles. For spectral resolutions of 50~m\AA \ and 100~m\AA, the spectral sampling at the chromospheric core are 25 m\AA \ and 50~m\AA, respectively. The photospheric line wings are covered with a coarse grid of points separated by 1~\AA \ in all cases.

\subsection{Non-LTE inversions}\label{sec:nlteinv}
 
We performed NLTE inversions of our simulated Stokes profiles, which were first convolved with a Gaussian
function of 50 FWHM, without noise. 

\begin{figure*}[]
      \centering
      	\resizebox{0.95\hsize}{!}{\includegraphics[trim= 0.00cm 0.00cm 2cm 0cm, clip]{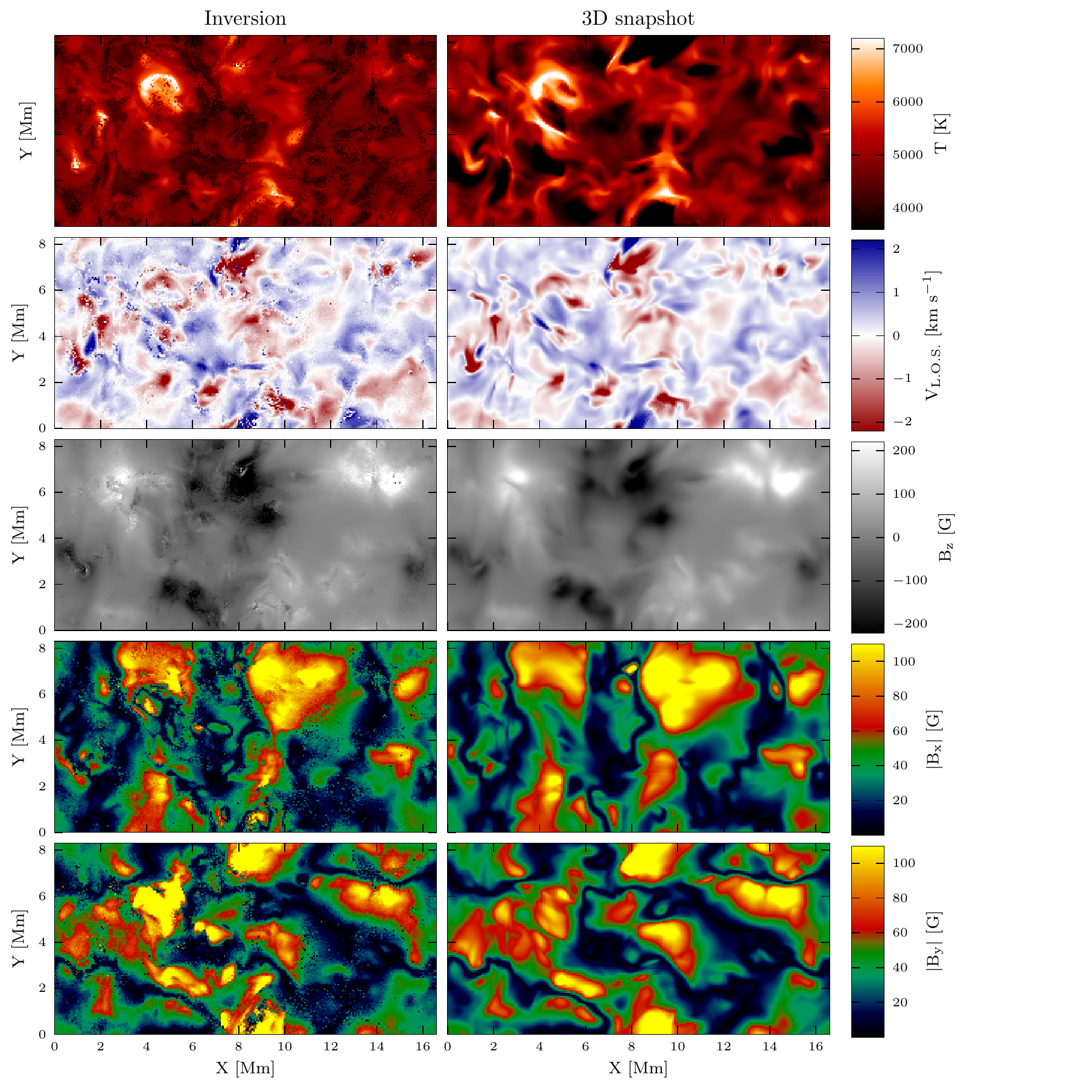}}
        \caption{Full field-of-view comparison between physical quantities in the chromosphere as determined from the NLTE inversion (\emph{left}) and the original 3D snapshot (\emph{right}). From top to bottom: temperature, line-of-sight velocity, $B_z$, $|B_x|$ and $|B_y|$. The line profiles were degraded with an instrumental profile of $\mathrm{FWHM}=50$~m\AA \ before carrying out the inversion. In the velocity panel, upflows have positive sign.}
        \label{fig:fullsize}
\end{figure*}

The results of these inversions are compared with the original atmosphere in Fig.~\ref{fig:fullsize}. Each quantity that is shown is the average over height between \ltau$=-4.5$ and the depth where the transition region starts at each pixel, on average \ltau$=-5.52$. The inversion code usually fails to reproduce the steep temperature gradient of the transition region, therefore it has been excluded from the depth average. In the 3D snapshot, the \ltau=$-4.5$ iso-surface corresponds on average to $\mathrm{z}=737~\pm~83$~km; the \ltau=$-5.52$ iso-surface is located on average at $\mathrm{z}=1340~\pm~365$~km. 
We stress that each column has its own run of $\tau_{5000}$ with height in the atmosphere. 
The derived quantities can therefore not be directly interpreted as the average over height in a horizontal slab.

The first row in Fig.~\ref{fig:fullsize} compares temperatures.
The inversion reproduces most of the high-temperature structure, but fails to reproduce the low temperatures. This failure is at least partly explained by the effect of 3D radiation, which is not taken into account in the inversion, because this assumes each pixel is a 1D plane-parallel atmosphere. Despite the failure to reproduce low temperatures, the average relative error in the inferred temperature (measured as the difference between model and inversion divided by the model) is only 6.7\%.  The Pearson correlation coefficient for the temperature is $r=0.51$. If we exclude cold areas ($T<3500$~K), the correlation factor becomes $r=0.67$.

The second row of Fig.~\ref{fig:fullsize} shows the line-of-sight velocity. The inversion is able to recover almost the entire structure originally present in the simulation, with a Pearson correlation
coefficient of $r=0.80$. The differences between the model and the inversion are mainly caused by inversion noise. Although in this paper we explore the inversion of one spectral line, it is still possible to retrieve information about gradients (mostly on the line-of-sight velocity and the magnetic field)
because these gradients are needed to reproduce the line asymmetries. The extreme case of a shock propagating through the atmosphere is discussed below in more detail.

The third row shows the vertical component of the magnetic field. The inversion reproduces the values in the atmosphere very well, with a Pearson correlation coefficient of $r=0.97$. The fourth and fifth row finally show the horizontal components of the magnetic field. The quality of the reconstruction is lower than for $B_z$, but still very good, with correlation coefficients of $r=0.90$ for $|$\bx$|$ and $0.78$ for $|$\by$|$. We note the presence of some artifacts in the inverted $|$\bx$|$ and $|$\by$|$ maps, such as in $|$\bx$(x=6.5,y=4.5)|$, where the inverted field is stronger than in the 3D snapshot. These artifacts appear because the inversion is computed here with one node in $|$\bx$|$ and one node in $|$\by$|$. The presence of strong photospheric fields can pollute the inverted values. Unfortunately, increasing the number of nodes also increases the inversion noise. 

\begin{figure*}[]
      \centering
      	\resizebox{\hsize}{!}{\includegraphics[trim= 0.00cm 0.00cm 0cm 0cm, clip]{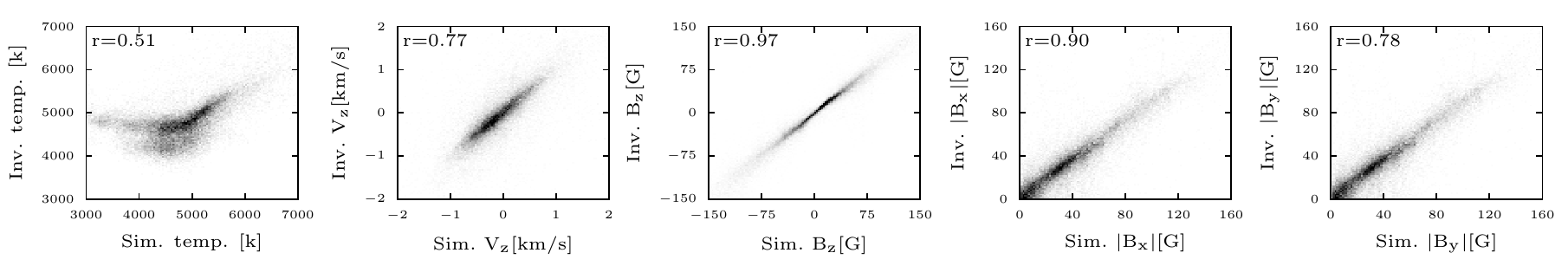}}
        \caption{Probability density plots of the inverted quantities vs. the 3D simulation quantities. From left to right: temperature, $V_{z}$, $B_z$, $|B_x|$ and $|B_y|$. The Pearson correlation coefficient (r) is indicated in the top-left corner of each panel.}
        \label{fig:scatter}
\end{figure*}

Figure~\ref{fig:scatter} shows probability density plots comparing the inverted quantities with the original data. All inverted quantities show a tight correlation with the original data, except for the temperature.

\begin{figure}[]
      \centering
      	\resizebox{\hsize}{!}{\includegraphics[trim= 0.00cm 0.00cm 0.cm 0.0cm, clip]{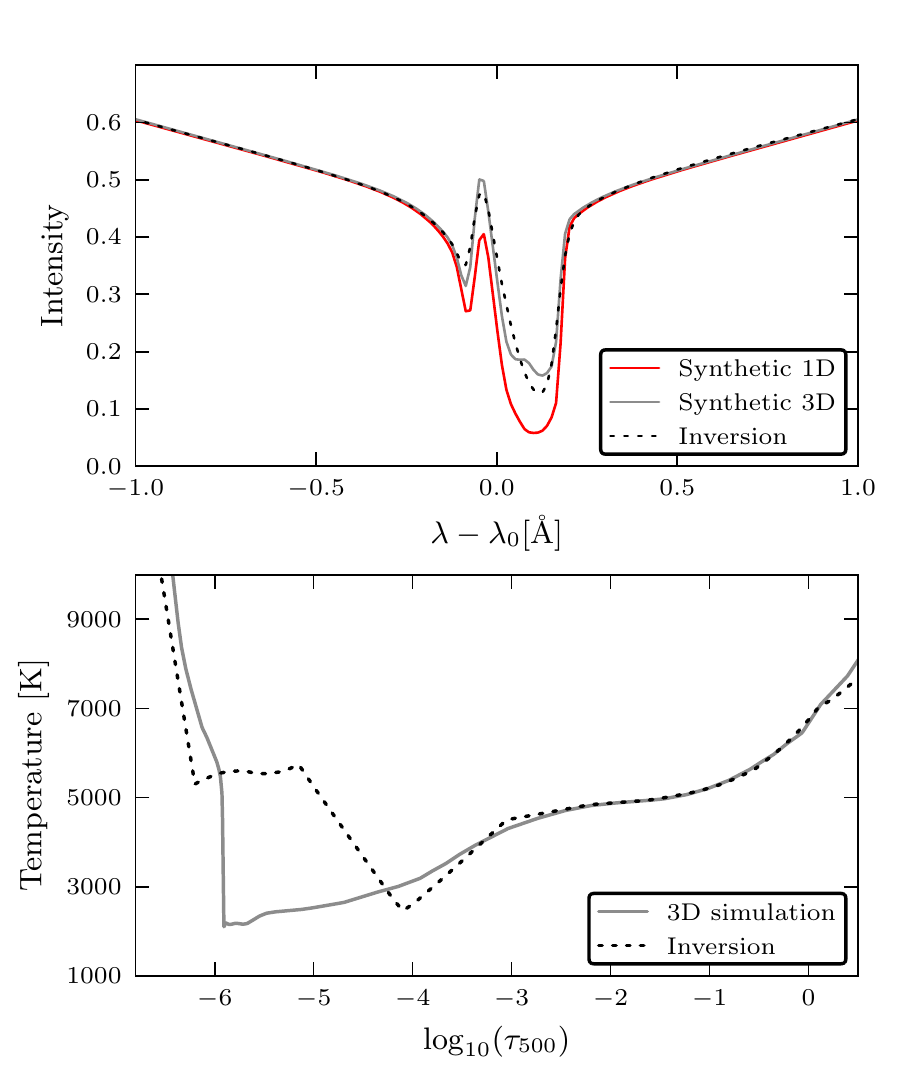}}
        \caption{\emph{Upper panel:} Vertically emerging intensity profile for a pixel with a low chromospheric temperature computed with \textsc{Multi3D} using 3D radiative transfer (solid-gray), the profile for the same pixel, but assuming 1D plane-parallel geometry (solid-red) and the best fit from an inversion (dashed). \emph{Lower panel}: The temperature structure for the same pixel (solid-gray) and the temperature structure inferred from the inversion (dashed).     
        \label{fig:darkregions}}
\end{figure}

The failure to reproduce low temperature areas is further illustrated in Fig.~\ref{fig:darkregions}. The top panel compares the emerging intensity profile for a pixel with a cold chromosphere (shown in the lower panel) computed in 3D and in 1D. In 1D, the line-core intensity is low, as a consequence of the low temperature. In 3D, however, the cold gas is illuminated from the sides, which increases the source function and hence the emerging intensity. The lower panel shows the temperature in the model atmosphere, and the temperature from the inversion. The inversion shows a much higher temperature (and hence source function) to reproduce the relatively high 3D line-core intensity. 

The presence of shocks in the atmosphere poses a particular problem for the inversion. Owing to the limited number of nodes, the inversion will often not be able to recover the true atmospheric structure. Instead it will try to find a much smoother structure that nevertheless approximates the emerging intensity.

\begin{figure}[]
      \centering
      	\resizebox{\hsize}{!}{\includegraphics[trim= 0.00cm 0.00cm 0.cm 0.0cm, clip]{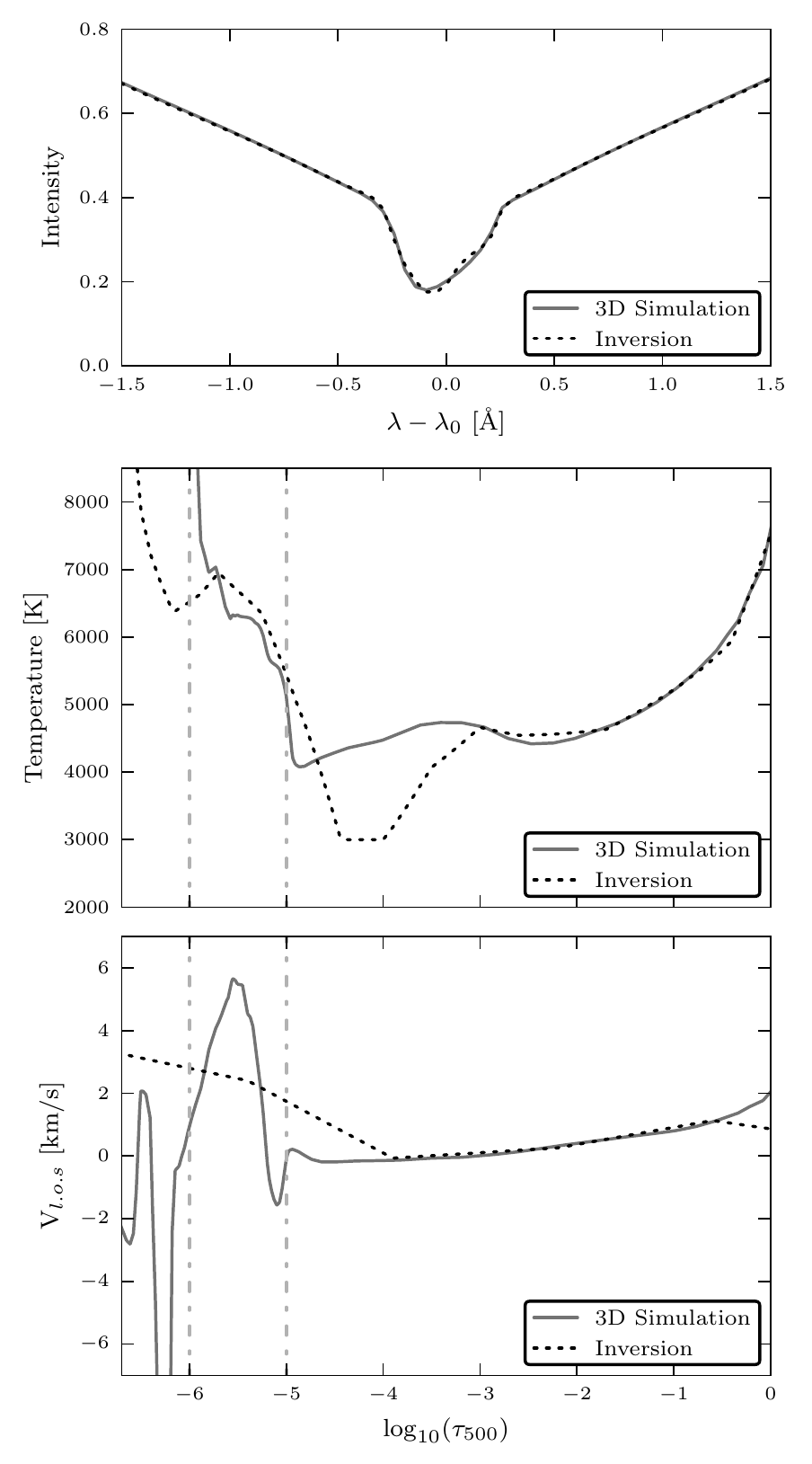}}
        \caption{\emph{Upper panel:} Intensity profile calculated from a column of the model where a shock is present (solid line). The dashed line corresponds to the inversion result. \emph{Middle panel: } Temperature stratification in the original model (solid) and the inversion result (dashed). \emph{Bottom:} Line-of-sight velocity stratification in the original model (solid) and the inversion result (dashed). The shock is located between the two gray dash-dotted lines.
        \label{fig:shock}}
\end{figure}

We illustrate this with an example from the 3D simulation in Fig.~\ref{fig:shock}: the shock is located between $\log_{10}(\tau_{500})=-5$ and $\log_{10}(\tau_{500})=-6$, as indicated by the increasing temperature and a spike in the vertical velocity. Together these lead to an asymmetric line core. The profile minimum is blueshifted with respect to the rest line-center frequency and the profile shows an inflection point at $\Delta \lambda=0.16$\,\AA. 

The profile-minimum intensity is well reproduced by the inversion, as is the profile-minimum Doppler shift, but the inflection point is not recovered. That the profile-minimum intensity is recovered means that the inferred temperature follows the true temperature in the shock region. The atmospheric velocity is not accurately recovered: the limited number of nodes forces a smooth inferred velocity field, which yields the same profile-minimum Doppler shift as the forward calculation, but has a lower maximum velocity. The inferred velocity has a smooth transition from~2 to~0~km\,s$^{-1}$ between $\log_{10}(\tau_{500})=-5$ and $\log_{10}(\tau_{500})=-4$, whereas the atmosphere has a much steeper velocity gradient. The lack of this steep gradient causes the lack of the inflection point in the inverted profile.

\begin{figure*}[!H]
      \centering
      	\resizebox{\hsize}{!}{\includegraphics[trim= 0.00cm 0.00cm 00.0cm 0.00cm, clip]{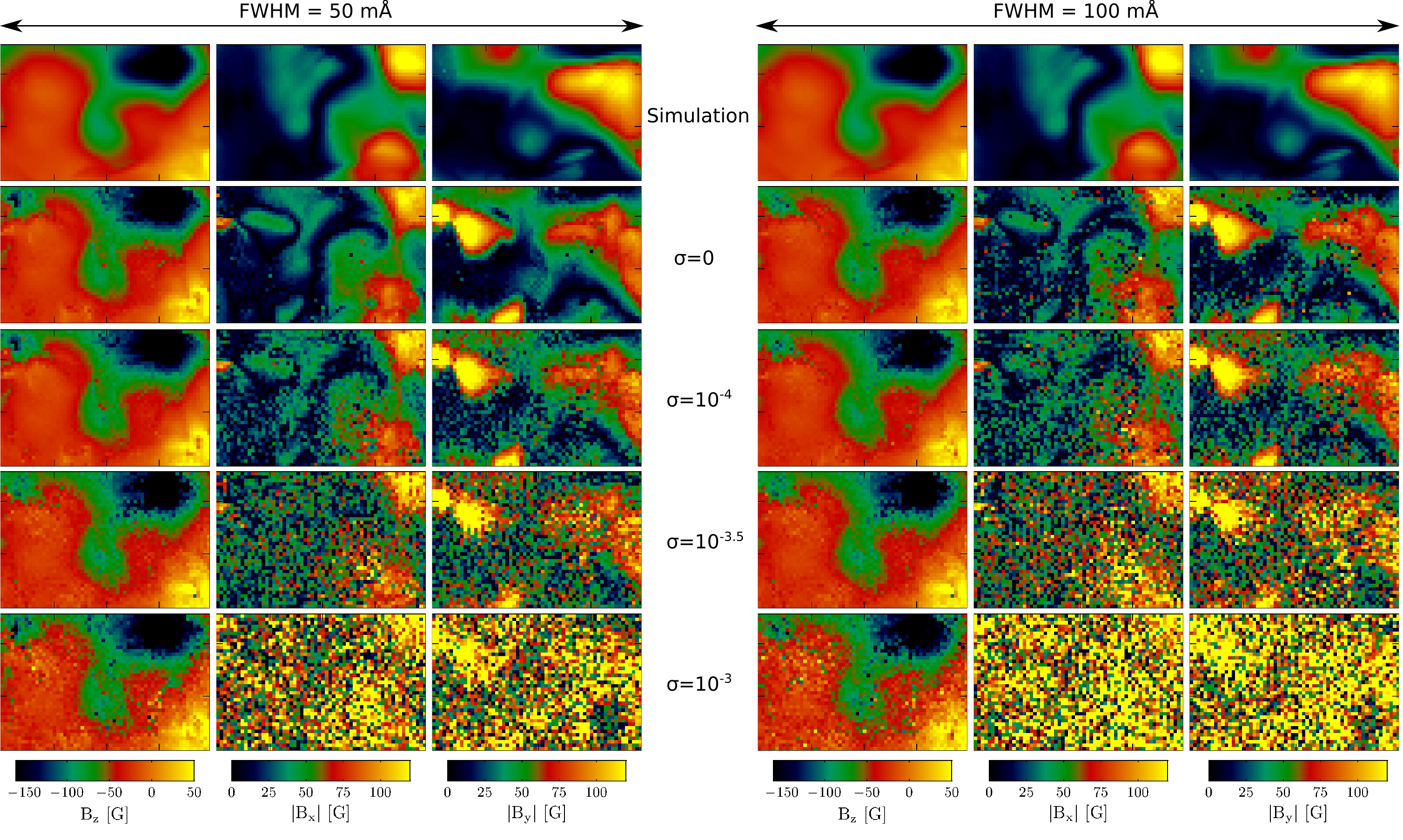}}

        \caption{Illustration of the effect of spectral resolution and photon noise on the quality of the inversions. The left-hand mosaic shows the results of the inversion after spectral degradation of the line profiles with a Gaussian instrumental profile with $\mathrm{FWHM}= 50$~m\AA, the right-hand panel after smearing with a profile with a $\mathrm{FWHM}$\ of 100~m\AA. Within each mosaic, from left to right, the panels show $B_z(x,y), |B_x(x,y)|, |B_y(x,y)|$. Rows, from top to bottom, show the magnetic field from the 3D snapshot, followed by the inversion results assuming random photon noise of $\sigma=0$, $\sigma=10^{-4}$, $\sigma=10^{-3.5}$ and  $\sigma=10^{-3}$. }
        \label{fig:bfield}
\end{figure*}

Fig.~\ref{fig:bfield} shows the effect of spectral resolution and photon noise on the quality of the inversions. Because of the large amount of computational work, we restricted the inversions to an area of 4$\times$2.5\ Mm$^2$ with its lower left corner at $(x,y)=(6,3)$\ Mm in Fig.~\ref{fig:fullsize}. We have introduced Gaussian noise of $\sigma=10^{-4}, 10^{-3.5}$ and $10^{-3}$. Comparison of the left-hand and right-hand mosaics shows that the quality of the inversion is more affected by noise levels than by the spectral degradation considered in our inversions. For a given noise level, the results obtained for a spectral resolution of 50~m\AA \ are slightly better than at 100~m\AA.

The longitudinal magnetic field (\bz) is the most accurately
retrieved parameter from the inversions. The Stokes~$V$ profiles
usually peak well above the $10^{-3}$ level. The reconstructed longitudinal field becomes increasingly noisy with increasing photon noise, but the overall structure  remains the same.

The maximum synthetic Stokes $Q$ and $U$ signal is typically only 5$\times$10$^{-4}$. Reconstruction of the horizontal components of the magnetic field is therefore much more sensitive to the noise level. At $10^{-4}$ noise, the inversion is able to recover most of the structure present in the atmosphere, more successfully when the $\mathrm{FWHM} = 50$~m\AA. When noise is increased to $\sigma=10^{-3.5}$, most of the details are lost, but areas with horizontal field larger than $\sim$100\ G are still discernible. Finally, when the noise is of the order of $10^{-3}$, the inversion is unable to recover the transverse magnetic field components.

\subsection{Velocity and magnetic field in LTE}

The intensity absorption profile of a given spectral line is given by
the expression \citep[see, e.g.][]{LDI92}
\begin{multline}	
	\kappa_I(\lambda, T, \vec B, v_{los})= \kappa_C(T) + {1 \over 2} 
	\eta_0(T, n_l, n_u)  \bigg{[}\phi_p(\lambda, \vec B, v_\text{los}) \sin2 \theta  \\
	+{\phi_b(\lambda, \vec B, v_\text{los}) + \phi_r (\lambda, \vec B, v_\text{los}) \over 2}(1+\cos2 \theta) \bigg{]} ,
	  \label{eq:absprofile}
\end{multline}
where $\kappa_C$ is the continuum absorption coefficient, $\eta_0$ is the line strength, $\phi_p$, $\phi_b$ and $\phi_r$ are the $\pi$, blue and red profile components, respectively.
Eq.~\ref{eq:absprofile} explicitly shows the dependencies on the temperature, $T$ 
(and other
thermodynamical parameters such as density or pressure), magnetic
field vector $\vec B$ and the line-of-sight velocity \vlos. For the continuum absorption coefficient $\kappa_C$ we have neglected the wavelength dependence because the continuum opacity varies with wavelength much more slowly than the line opacity. The atomic level populations of the lower and upper levels of the transition considered, $n_l$ and $n_u$, obviously depend on $T$ as well.
Even in conditions far from LTE, the atomic populations are only weakly dependent on $\vec B$ and \vlos, unless the velocity gradients are very steep.

In Eq.~\ref{eq:absprofile} there is a convenient separation of the atmospheric parameters (the unknowns of our inversions). The dependence on the thermodynamical parameters $T$ is contained exclusively in $\kappa_C$ and $\eta_0$, whereas the dependencies on the magnetic field $\vec B$ and the velocity \vlos \  are in the term enclosed in square brackets. We can then see that the wavelength dependence of the profile (which is contained in the terms within the brackets) depends only on the values of $\vec B$ and \vlos . The same happens with the equivalent absorption profiles for the rest of the Stokes parameters $\kappa_Q$, $\kappa_U$, $\kappa_V$, as well as the emission and the anomalous dispersion profiles. They can all be separated into a term containing the atomic populations and the thermodynamic properties and another term containing the wavelength profiles with the dependence on $\vec B$ and $v_\text{los}$.
\begin{figure*}[]
      \centering
        \resizebox{\hsize}{!}{\includegraphics[trim=0cm 0.0cm 1.4cm 0cm, clip=true]{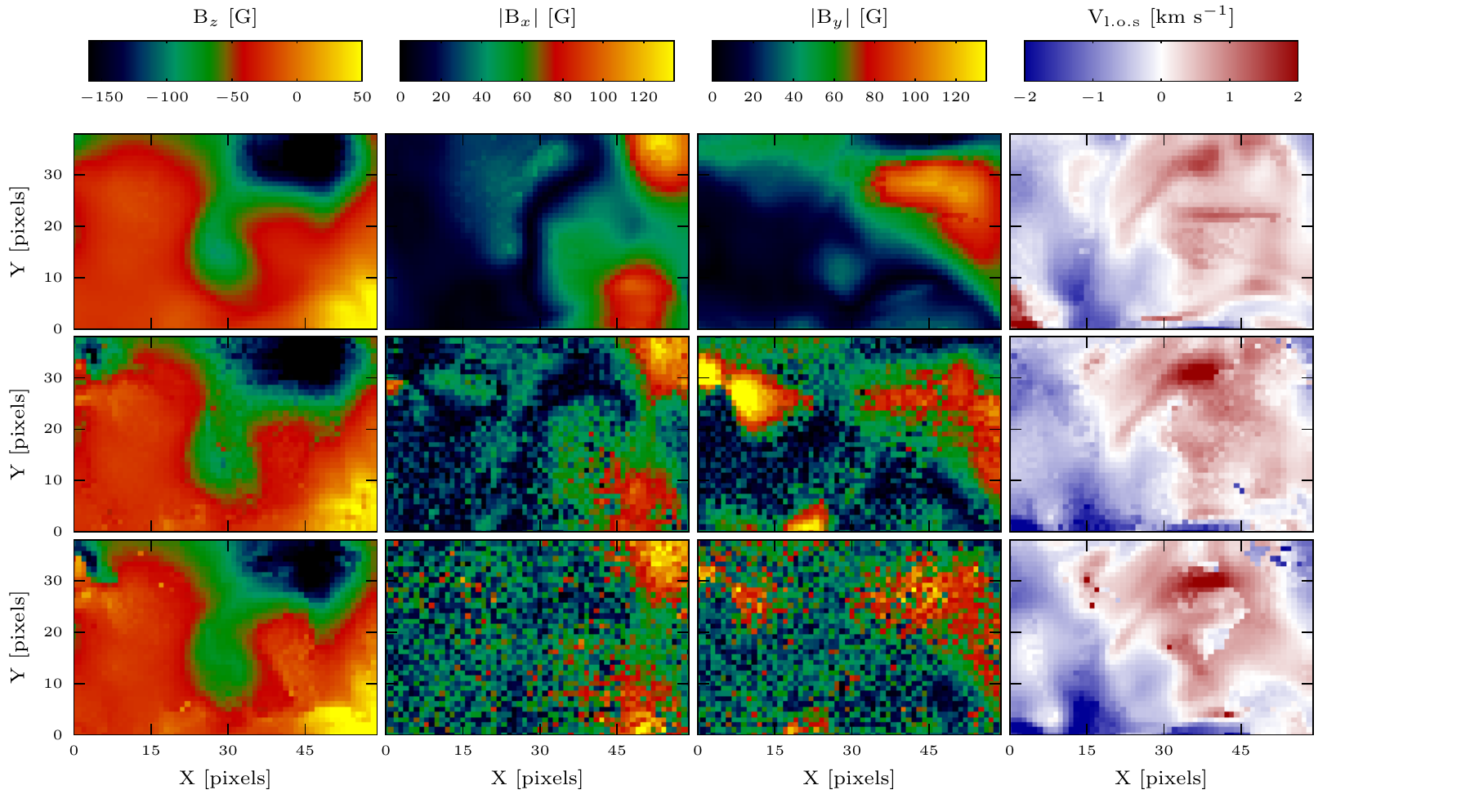}}
        \caption{Results from an inversion carried out in LTE. The top row illustrates
          quantities from the 3D snapshot, the middle row the results of the NLTE inversion and the bottom row the
          results of the LTE inversion. In the velocity panel, upflows have positive sign. The inversions were computed assuming a spectral resolution of 50 m\AA \ and random noise with $\sigma=10^{-4}$.}
        \label{lteres}
\end{figure*}

This separation of variables is fortunate because if we are not interested in the temperature but only in the magnetic field and velocity, it is possible to devise a much simpler inversion for these quantities. This simpler inversion would not require the solution of an NLTE problem, and the computational demands are much lower.

Let us consider now one such simplistic procedure, namely one that
fits the observed profiles computing the atomic populations in
LTE. With the inversion we obtain a run of thermodynamic parameters
$T^*$ which is, in general, different from the real $T$ in the solar
atmosphere. From that $T^*$ we compute LTE populations $n^*_l$ and
$n^*_u$. 

When the observations are fitted by the procedure, the source function
and the opacity are adequately reproduced. Therefore, the populations
obtained are approximately correct ($n^*_l \simeq n_l$ and $n^*_u
\simeq n_u$), even though the temperatures needed to yield those
values are drastically different in LTE and NLTE. The wavelength
dependence of the observed spectral line depends entirely on $\vec B$
and \vlos, whose effects on the profile are the same regardless of
whether the populations are computed in LTE or NLTE. In summary, with
our simplistic LTE fitting procedure we would recover $\vec B$ and
\vlos with the same degree of accuracy as with an NLTE inversion, and only $T$
would be wrong.

With the same reasoning, we argue that questioning the validity of the
assumptions that we have made on the radiative transfer (such as
hydrostatic equilibrium, 1.5D atmosphere or the assumption of statistical equilibrium)
would only be relevant in determining $T$ but not $\vec B$ or
$v_{los}$.

This approach is actually similar to what is often performed in
Milne-Eddington inversions for the photosphere, in which only $\vec B$ and
$v_{los}$ are interpreted as physical parameters, even though several
other parameters are involved in the fit.

Figure~\ref{lteres} depicts the result of this LTE inversion  of the synthetic Stokes profiles to determine $\vec{B}$ and \vlos \ in the same region of the simulation as for Fig.~\ref{fig:bfield}. The inversions in Fig.~\ref{lteres} correspond to $\sigma=10^{-4}$ and a spectral resolution of 50~m\AA. Comparison of the inferred \bz \ and \vlos \ shows that the inversion performs reasonably well in reproducing these quantities. The noise comes primarily from inversions that failed to converge. The Pearson correlation coefficients are $r=0.93$ for \bz \ and $r=0.84$ for \vlos.

For \bx \ and \by \ the results appear more noisy than for \bz, but the inversion more or less correctly reproduces the patches of strong field on the right side of the panel. The inversion erroneously infers patches of horizontal field where there are none in the atmosphere, such as at $(x,y) = (12,28)$. The Pearson correlation coefficients are $r=0.90$ and $r=0.69$ for \bx \ and \by, respectively. The lower quality of the inversion compared to the \bz \ case is due to the Stokes~$Q$ and $U$ signals being weaker than Stokes~$V$.

In a realistic case, the data would be dominated by the
observational noise, except in specific cases such as strong
fields in the vicinity of sunspots.

\section{Discussion and conclusions}\label{conclusions}
We have computed full Stokes synthetic line profiles using a 3D radiation-MHD simulation of the outer atmosphere of the Sun. A full 3D solution  of the NLTE radiative transfer problem was performed using \textsc{Multi3D}.  The width of the core of the synthetic \ion{Ca}{ii} intensity profiles is approximately a factor 2.5 lower than observed. This would lead to artificially strong Stokes $Q$, $U$ and $V$ signal. To remedy this, external micro-turbulence was added, which brought the synthetic line widths closer to the observed ones.

The resulting line profiles were degraded by various amounts of spectral smearing with various levels of noise added. They were then inverted using the inversion code \textsc{Nicole}, which solves the NLTE radiative transfer problem assuming plane-parallel geometry for each pixel (1.5D). The inversions were performed assuming that the polarization is produced by the Zeeman effect only. This assumption is reasonable for solar regions with moderate to strong activity, which is the most likely target for an application of \textsc{Nicole}.

We find here that the NLTE inversion can reliably recover the original line-of-sight magnetic field and velocity, with the noise levels and spectral resolutions considered. Temperatures are less reliably inverted. High-temperature areas are well reproduced, but the inversion is less sensitive to cold areas because it cannot properly treat the 3D effect caused by incoming radiation from hotter neighboring pixels.

The quality of the NLTE inversion for the horizontal field strongly depends on the amount of noise present in the Stokes profiles, and less so on the spectral resolution. Only with a noise level below $\sigma=10^{-3.5}$, is the inversion able to recover the features present in the model atmosphere.

We also tested the reliability of a simpler and faster LTE inversion, which can recover the line-of-sight velocity and the magnetic field, but not the temperature. We showed that this inversion recovers \bz\ and \vlos\ quite well, but yields less reliable results for the horizontal components of the magnetic field. Nevertheless, LTE inversion is a viable and fast strategy if one is interested in the vertical velocity and magnetic field only.

Our results strongly suggest that to be able to detect and invert Stokes $Q$ and $U$ signals in the quiet-Sun, observations should have a noise level better than $\sigma = 10^{-3.5}$. 
Instrumental filter widths below 50~m\AA\ do not improve the quality of the inversion.

Our work leads to a better understanding of the information contained in chromospheric \ion{Ca}{ii} observations and what can be learned from their inversions. We showed that a NLTE inversion can reliably recover the vertical magnetic field and velocity at noise levels of $\sigma = 10^{-3}$ from a radiation-MHD simulation of quiet Sun, and that the horizontal components can be recovered if the noise level is below $\sigma=10^{-3.5}$. Our results also strongly suggest that if the inversion techniques are applied to observations of more active regions on the Sun, which produce stronger polarization signals, the inferred quantities will be fairly accurate.

\begin{acknowledgements}
HSN gratefully acknowledges financial support by the Spanish Ministry of Science and Innovation through project AYA2010-18029 (Solar Magnetism and Astrophysical Spectropolarimetry) and CONSOLIDER INGENIO CSD2009-00038 (Molecular Astrophysics). This research project has been supported by a Marie Curie Early Stage Research Training Fellowship of the European Community's Sixth Framework Programme under contract number MEST-CT-2005-020395: The USO-SP International School for Solar Physics. This research was supported by the Research Council of Norway through
 the grant ``Solar Atmospheric Modelling'' and through grants of computing time from the Programme for Supercomputing. JdlCR gratefully acknowledges financial support by the European Commission through the SOLAIRE Network (MTRN-CT-2006-035484).  JL acknowledges support from the Netherlands Organization for Scientific Research (NWO).
 \end{acknowledgements}

\bibliography{delacruz}

\end{document}